\documentclass[prd,aps,eqsecnum,floatfix,nofootinbib,preprint,tightenlines,superscriptaddress]{revtex4}
\usepackage{dcolumn}
\usepackage{amsmath}
\usepackage{latexsym}
\usepackage{amssymb}
\usepackage{graphicx}
\usepackage{bm}
\usepackage{colordvi}
\usepackage{xcolor}

\def\a{\alpha}

\def\c{\chi}
\def\g{\gamma}

\def\j{\psi}
\def\bj{\bar\psi}

\def\m{\mu}
\def\n{\nu}

\def\p{\pi}                     

\def\x{\xi}

\def\O{\Omega}
\def\P{\Pi}

\def\S{\Sigma}

\def\cd{{\cal D}}

\def\co{{\cal O}}

\def\qhat{{\hat q}}

\def\bx{{\bf x}}

\def\svev#1{\left\langle #1\right\rangle}       

\def\tr{{\rm tr}\,}

\long \def \blockcomment #1\endcomment{}

\def\Eq#1{Eq.~(\ref{#1})}

\def\bra#1{\mathinner{\langle{#1}|}}
\def\ket#1{\mathinner{|{#1}\rangle}}

\def\SU{\text{SU}}
\def\SO{\text{SO}}
\def\PiLR{\Pi_{LR}}
\def\CLR{ C_{LR}}

\newcommand{\bee}{\begin{equation}}
\newcommand{\ee}{\end{equation}}
\newcommand{\beea}{\begin{eqnarray}}
\newcommand{\eea}{\end{eqnarray}}

\newcommand{\chidof}{\ensuremath{\mbox{$\chi^2/\text{d.o.f.}$}}}

\begin{document}
\title{
Radiative contribution to the effective potential in composite Higgs models from lattice gauge theory}
\author{Thomas DeGrand}
\affiliation{Department of Physics,
University of Colorado, Boulder, CO 80309, USA}
\author{Maarten Golterman}
\affiliation{Department of Physics and Astronomy, San Francisco State University, San Francisco,
CA 94132, USA}
\author{William I.~Jay}
\affiliation{Department of Physics,
University of Colorado, Boulder, CO 80309, USA}
\author{Ethan T.~Neil}
\affiliation{Department of Physics,
University of Colorado, Boulder, CO 80309, USA}
\affiliation{RIKEN-BNL Research Center, Brookhaven National Laboratory,
Upton, NY 11973, USA}
\author{Yigal Shamir}
\author{Benjamin Svetitsky}
\affiliation{Raymond and Beverly Sackler School of Physics and Astronomy,
Tel~Aviv University, 69978 Tel~Aviv, Israel}

\begin{abstract}

We develop methods to calculate the electroweak gauge boson contribution to the effective Higgs potential in the context of composite Higgs models, using lattice gauge theory.  The calculation is analogous to that of the electromagnetic mass splitting of the pion multiplet in QCD.  We discuss technical details of carrying out this calculation, including modeling of the momentum and fermion-mass dependence of the underlying current-current correlation function; direct integration of the correlation function over momentum; and  fits based on the minimal-hadron approximation.  We show results of a numerical study using valence overlap fermions, carried out in an SU(4) gauge theory with two flavors of Dirac fermions in the two-index antisymmetric representation.

\end{abstract}

\maketitle

\section{Introduction}
Composite Higgs theories \cite{Georgi:1984af,Dugan:1984hq} have often been written down as effective field theories, typically non-linear sigma models \cite{Contino:2010rs,Bellazzini:2014yua,Panico:2015jxa}.
In one approach, the sigma model describes a set of exactly massless Nambu--Goldstone bosons that live in a coset manifold $G/H$.
This set contains the Higgs multiplet of the Standard Model (SM).
The Higgs potential then comes mainly from coupling to the electroweak
gauge bosons and to the top quark; this potential should induce the Higgs phenomenon of the SM.
If the SM's gauge symmetries are a subgroup
of the unbroken symmetry $H$, then the coupling to the SM's gauge bosons will
induce a positive curvature at the origin of the Higgs potential, a phenomenon known as
vacuum alignment~\cite{Peskin:1980gc}.
A negative curvature, required in order to induce
electroweak symmetry breaking, can then arise only from the top quark's contribution.

The sigma model is only an effective low-energy description with the correct symmetry
properties.  A fuller understanding of the dynamics requires an underlying, ultraviolet-complete
theory that gives rise to the non-linear sigma model at low energies.  Several such proposals now exist in the literature \cite{Barnard:2013zea,Ferretti:2014qta,Vecchi:2015fma,Ma:2015gra}, including an effort to catalogue ultraviolet completions
\cite{Ferretti:2013kya,Ferretti:2016upr} that accommodate both a composite Higgs and a partially composite top quark~\cite{Kaplan:1991dc}.

Within such an ultraviolet completion, lattice simulations \cite{DeGrand:2015zxa} can be used to determine low-energy constants of the theory, including the radiative contributions of the SM to the Higgs potential.  In this paper, we present a lattice calculation for the gauge bosons' part of the Higgs potential.
This is given by%
\footnote{
This is the expression in the case of a coset of the form ${\rm SU}(N)/{\rm SO}(N)$,
which is relevant to the model considered here.   For other cosets the general form is similar.
The top-sector contribution to the effective potential
  for the model of Ref.~\cite{Ferretti:2014qta} was worked out
  in Ref.~\cite{Golterman:2015zwa}.
  A lattice calculation of this contribution is more challenging.
}
 \cite{Contino:2010rs,Golterman:2014yha,Golterman:2015zwa}

\bee
V_{\rm eff} = C_{LR}\sum_Q\tr\left(Q\S Q^*\S^*\right)\ ,
\label{LEC}
\ee
where $\Sigma$ is the non-linear field representing
the multiplet of pseudo-Nambu--Goldstone bosons that contains the composite Higgs boson.
The sum over $Q$ runs over the $SU(2)_L$ generators $gT^a_L$
and the hypercharge generator $g'Y$, with $g$ and $g'$ the corresponding
coupling constants.  In accordance with vacuum alignment,
the low-energy constant $C_{LR}$ is positive
\cite{Witten:1983ut,Golterman:2014yha}, and the minimum of $V_{\rm eff}$,
given by $-C_{LR}(3g^2+g'^2)$, is attained at $\langle\S\rangle={\bf 1}$.

In principle, $\CLR$ can be determined in a lattice calculation
where the gauge bosons of both the new strong interaction and of
the electroweak interactions are present as dynamical degrees of freedom.
A more economical approach, which avoids the introduction of electroweak
gauge bosons into the lattice simulation, is to obtain $\CLR$ in terms of a correlation function of the ultraviolet theory,
\bee
  \CLR = \int_0^\infty dq^2 q^2\, \PiLR(q^2).
  \label{eq:CLR1dintegral}
\ee
Here $\PiLR(q^2)$ is the transverse part of the current--current correlation function,
\beea
\frac12\delta_{ab}\Pi_{\mu\nu}(q) &=& -\int {d^4 x} \,e^{iqx}\, \langle J_{\mu a}^L(x) J_{\nu b}^R(0)\rangle,
\label{PiLRdefn}
\\[2pt]
  \Pi_{\mu\nu}(q) &= & (q^2 \delta_{\mu\nu} - q_\mu q_\nu)\Pi_{LR}(q^2) + q_\mu q_\nu\Pi'(q^2) .
  \label{PiLRdecomp}
\eea
The chiral currents are
\beea
J^L_{\mu a}&=&\bar\psi\gamma_\mu(1-\gamma_5){T_a}\psi
= V_{\mu a}-A_{\mu a}\,,\nonumber\\
J^R_{\mu a}&=&\bar\psi\gamma_\mu(1+\gamma_5){T_a}\psi
= V_{\mu a}+A_{\mu a}\,,
\eea
where $T_a$ are the isospin generators.
In the chiral limit, where the low-energy constant in \Eq{LEC} is defined, the current correlator (\ref{PiLRdefn}) is automatically transverse.

Our task is related to a classic problem in hadronic physics.
In quantum chromodynamics with two flavors, the pions are pseudo-Nambu--Goldstone bosons, massless in the chiral limit.
The entire isotriplet gets most of its mass from chiral symmetry breaking by the quark masses;
the charged pion gets an additional contribution from coupling to electromagnetism, lifting it above the neutral pion.
Defining $\Delta m_\pi^2 \equiv m^2_{\pi^{\pm}}-m^2_{\pi^0}$, we quote
the sum rule derived long ago by Das et~al.~\cite{Das:1967it},
\begin{equation}
\Delta m_\pi^2=
-\frac{3\alpha}{4\pi f_\pi^2}\int_0^\infty dq^2
\left.q^2\Pi_{LR}(q^2)\right|_{m_q=0}.
\label{eq:Das}
\end{equation}
This has served as the basis of lattice calculations of the pion mass difference that have been quite successful~\cite{Shintani:2008qe,Shintani:2008ga}.%
\footnote{For an early lattice discussion, see Ref.~\cite{Gupta:1984tb}.}

We present here a  lattice calculation of the quantity $\CLR$ in the context of composite Higgs theories.
We study an SU(4) gauge theory with $N_f = 2$ Dirac fermions in the sextet representation of SU(4), which is the antisymmetric two-index representation---a real representation.  We have studied this theory before \cite{DeGrand:2015lna}, focusing on the phase diagram of the lattice theory and its particle spectrum.  This theory is the first step towards a lattice simulation of the UV-complete composite Higgs model proposed in \cite{Ferretti:2014qta}, in which 5 flavors of Majorana fermions in the sextet representation of SU(4) give rise to a composite Higgs within an SU$(5)/$SO(5) coset.  The 5-Majorana theory can be studied on the lattice, but it requires the use of methods such as rational HMC (RHMC), which increase the computational cost compared to the theory with two Dirac fermions.%
\footnote{To construct SU(4) baryons that can couple linearly to the top quark, one must also add Dirac fermions in the fundamental (quartet) representation of SU(4).
There are 3 flavors of them, making the baryons into an anti-triplet of ordinary color.
The properties of this model have been studied at length in Refs.~\cite{Ferretti:2014qta,Golterman:2015zwa}.
We do not consider these in the present paper.}

As noted above, we require $\CLR$ in the massless limit.
In fact, $\PiLR(q^2)$ is an order parameter for the spontaneous breaking of chiral symmetry.
Since we use Wilson--clover fermions in the action that generates gauge field configurations, chiral symmetry is not restored in the limit where the pion%
\footnote{We use henceforth QCD terminology for the meson spectrum of our SU(4) gauge theory.}
 becomes massless (as long as the lattice spacing is nonzero).
 Thus we are led to use overlap fermions \cite{Neuberger:1997fp,Neuberger:1998my}, with exact chiral symmetry, to define correlation functions calculated on the gauge field ensembles.
In this mixed-action theory we can vary the valence (overlap) quark mass while keeping the dynamical (Wilson--clover) mass fixed, and approach a chiral limit in the former.%
\footnote{Previous calculations of $\PiLR(q)$ in QCD and beyond \cite{Shintani:2008qe,Boyle:2009xi,Appelquist:2010xv}, have also used overlap (or domain-wall) fermions.}

$\PiLR(q^2)$ may be modeled by
retaining only the pion and $a_1$ poles in the axial channel,
and the $\rho$ meson's pole in the vector channel.
This is the minimal-hadron approximation%
\footnote{We note that the location of the pole at $q^2=0$ does not depend on the chiral limit; see Appendix \ref{sec:pole}.}
 (MHA),
\bee
  \PiLR(q^2) \approx
  \frac{f_\p^2}{q^2} + \frac{f_{a_1}^2}{q^2+m_{a_1}^2}
  - \frac{f_\rho^2}{q^2+m_\rho^2} \,.
\label{MHA}
\ee
In the chiral limit, one can use the Weinberg sum rules
\cite{Weinberg:1967kj} to eliminate $f_\rho$ and $f_{a_1}$,
giving
\bee
  \PiLR(q^2) \approx \frac{f_\p^2}{q^2}
  \frac{m_{a_1}^2 m_\rho^2}{(q^2+m_{a_1}^2)(q^2+m_\rho^2)} \,,
\label{PiMHA}
\ee
and thus
\bee
  \CLR \approx f_\p^2\, \frac{m_{a_1}^2 m_\rho^2}{m_{a_1}^2-m_\rho^2}\,
  \log\left(\frac{m_{a_1}^2}{ m_\rho^2}\right) \,.
\label{CMHA}
\ee
In the QCD case, the MHA misses the experimental value of $\Delta m_\pi^2$ by 25\%.
Our numerical calculation of $\PiLR$  enables us to test the approximation (\ref{MHA}) directly in the present theory.

This paper deals more with technique than with results.
Hence, while we present numerical data for two different lattice actions, we do not take either the continuum limit or the chiral limit for the dynamical
(``sea'') fermions.
Indeed, the two ensembles that we study produce values of $\CLR$ that are in disagreement.
As we have stated above, the theory studied here is not quite an actual model for a composite Higgs boson.
This is why we stop short of an extrapolation to the continuum limit, which would involve considerable additional computation for a theory that is, after all, not of direct physical interest.

Our paper is organized as follows.  In Sec.~\ref{sec:action} we present the technical aspects of our lattice simulations.
We specify the lattice action with which we generated configurations, and we describe the measurement of $\PiLR(q^2)$, defined via overlap valence fermions, on these configurations.  We then proceed to our numerical work.
We generated two ensembles with different lattice actions; we present the ensembles and their particle spectra in Sec.~\ref{sec:spectra}.
Once $\PiLR(q^2)$ is in hand, the integration in \Eq{eq:CLR1dintegral} is straightforward, except for handling the pole in $\PiLR$ at $q=0$.
We show one way to do this in Sec.~\ref{sec:CLR}.  In Sec.~\ref{sec:MHA}, we demonstrate an alternative method for calculation of
$\CLR$, fitting $\PiLR(q^2)$ to a rational function inspired by the MHA.  We find that the two methods for extracting $\CLR$ produce results that agree well, for each ensemble.
Our conclusions are in Sec.~\ref{sec:conclusions}.

We give some details about the lattice action in Appendix \ref{sec:NDS} and review some current algebra in Appendix \ref{sec:pole}.  In Appendix \ref{sec:ChiPT} we describe an attempt to reconcile our two ensembles using chiral perturbation theory.

\section{Lattice action and overlap valence fermions\label{sec:action}}

\subsection{Wilson--clover action with dislocation suppression}

We generate ensembles of gauge configurations with the lattice action,
\bee
S=S_{\text{plaq}}+S_{{f}}+S_{\text{NDS}}.
\ee
Here $S_{\text{plaq}}$ is the usual plaquette action for the fundamental gauge fields $U_{x\mu}$.
The fermion action $S_{{f}}$ is comprised of the conventional Wilson hopping term
and a clover term \cite{Sheikholeslami:1985ij}, where the gauge connection in both
terms is constructed from $U_{x\mu}$ in two steps: normalized hypercubic (nHYP)
smearing \cite{Hasenfratz:2001hp,Hasenfratz:2007rf,DeGrand:2012qa} followed by
promotion to the sextet representation.
Finally the term $S_{\text{NDS}}$ is a pure gauge term designed to suppress
dislocations in the dynamical gauge field, in order to help eliminate spikes in the fermion force and to ease the calculation of the overlap operator (see Appendix~\ref{sec:NDS} and Ref.~\cite{DeGrand:2014rwa}).

The action is identical to that used in our earlier study \cite{DeGrand:2015lna},
except for the introduction of the NDS term.
The gauge coupling is set by the coefficients $\beta$ and $\gamma$ of the
plaquette and NDS terms: at tree level,
\bee
\frac1{g_0^2}=\frac{\beta}{2N_c}+\frac{\gamma}{N_c}\left(\frac{\alpha_1}3+\alpha_2+\alpha_3\right),
\label{eq:g02}
\ee
where $\alpha_i$ are the smearing parameters and we have adopted a
common $\gamma$ for the three levels of smearing.
Our smearing parameters took the values $
(\alpha_1,\alpha_2,\alpha_3)=(0.75,0.6,0.3).
$

The theory contains two degenerate flavors of sea quarks, whose common bare mass is introduced via the hopping
parameter $\kappa=(2m_0 a+8)^{-1}$.
As is appropriate for nHYP smearing~\cite{Shamir:2010cq},  the clover coefficient
is set to its tree level value, $c_{\text{SW}}=1$.
Since the sextet is a real representation of the gauge group, the global
symmetry of the continuum theory is SU(4), which breaks spontaneously to
 SO(4); in the lattice theory, on the other hand, the Wilson term breaks the SU(4) explicitly to SO(4).
As noted in Ref.~\cite{DeGrand:2015lna}, the nHYP smearing procedure
introduces a small violation of the reality of the sextet gauge links.
This is a lattice artifact which breaks the degeneracies expected by virtue of the SO(4) symmetry.
We leave this uncorrected in the generation of gauge configurations, but we
 do apply a correction when we use the gauge configurations to calculate fermionic observables.
Thus the correlation functions of fermionic operators indeed satisfy the SO(4) symmetry.
Like other differences between sea quarks and valence quarks, the effects of
 this correction will disappear with the lattice spacing.

\subsection{Overlap operator}

For the valence quarks and their currents, we use an overlap operator.
Our implementation is as described in
 Refs.~\cite{DeGrand:2000tf,DeGrand:2004nq,DeGrand:2006ws,DeGrand:2007tm,DeGrand:2006nv}. The
 massless overlap operator is defined as
\bee
D(0)=R_0\left(1+ \frac{d(-R_0)}{\sqrt{d^\dagger(-R_0)d(-R_0)}}\right).
\ee
Here $d(m)=d+m$, where $d$ is a massless Wilson--Dirac operator on the lattice.
We choose one with nearest- and next-nearest-neighbor interactions,
plus a clover term; we set  $R_0=1.2$.
Adding a valence fermion mass $m_v$ gives the operator
\bee
D(m_v)=\left(1-\frac{m_v}{2R_0}\right)D(0)+m_v.
\ee
We evaluate the correlation function in $\PiLR$ using improved currents.
The vector current is
\bee
V_{\mu a}= \bar q \gamma_\mu T_a\left(1-\frac{aD(0)}{2R_0}\right) q ,
\label{eq:Vmu}
\ee
while the axial current is
\bee
A_{\mu a}= \bar q \gamma_\mu\gamma_5 T_a\left(1-\frac{aD(0)}{2R_0}\right) q.
\label{eq:Amu}
\ee
These currents are not conserved, and so the correlator of each current has a quadratically divergent contact term.
Thanks to a chiral Ward identity, the quadratic divergence cancels in the vector--axial difference.

In the actual computation of the correlator, we move the overlap operator out of the currents and into the quark propagator.
We replace the currents (\ref{eq:Vmu})--(\ref{eq:Amu}) by point currents and the quark propagator $D^{-1}$ by the ``shifted'' propagator \cite{Capitani:1999uz,DeGrand:2003in}
\bee
\hat D^{-1}(m_v) = \frac{1}{ 1-m_v/(2R_0)}\left[D^{-1}(m_v) - \frac{1}{2R_0}\right] .
\label{eq:SUBPROP}
\ee

We compute eigenvalues of the squared Hermitian Dirac operator
$D^\dagger D$ with the {\tt Primme} package \cite{primme}; these are used to precondition the calculation of propagators.

\subsection{Extraction of $\PiLR$ \label{subsec:qhat}}

As noted in the introduction, $\PiLR(q^2)$ is an order parameter for chiral symmetry.
Thanks to our use of overlap fermions, this remains true on the lattice, with the currents defined in the preceding subsection.
In analogy with \Eq{PiLRdefn}, we define the lattice vacuum polarization via
\bee
\frac12\delta_{ab}\Pi_{\mu\nu}^{\textrm{lat}}(q) = -\sum_x \,e^{iqx}\, \langle J_{\mu a}^L(x) J_{\nu b}^R(0)\rangle\ .
\label{eq:latPiLR}
\ee
We would like to decompose this as in \Eq{PiLRdecomp}.
Lattice artifacts, however, make the structure of
$\Pi_{\mu\nu}^{\textrm{lat}}(q)$ more complex,
\bee
\Pi_{\mu\nu}^{\textrm{lat}}(q) = P^T_{\mu\nu}(q) \PiLR(q) + P^L_{\mu\nu}(q) \Pi'(q) + \cdots
\label{eq:dirt}
\ee
where $P^T(q)$ and $P^L(q)$ are lattice analogs of the transverse and longitudinal projectors (see below).
The dots represent
additional terms that are proportional to higher powers of $q_\mu$ (and of the lattice spacing) and that break rotational invariance \cite{Aubin:2015rzx}.

We deal with these lattice artifacts empirically, following the method of Refs.~\cite{Shintani:2008qe,DeGrand:2010tm}.
For each lattice momentum $q$, we do a fit to $\Pi^{\textrm{lat}}_{\mu\nu}(q)$, treating $\PiLR$ and $\Pi'$ as fit parameters.
This means that we form the $\chi^2$ function for each individual momentum mode $q$,
\bee
\chi^2_q=\sum_{\mu\nu}\left[\Pi^{\textrm{lat}}_{\mu\nu}(q)-P^T_{\mu\nu}(q)\PiLR(q)-P^L_{\mu\nu}(q)\Pi'(q)\right]^2,
\ee
taking the sixteen $\Pi^{\textrm{lat}}_{\mu\nu}$ correlators as the quantities to be fit.
Here
\beea
P^T_{\mu\nu}(q)&=& \qhat^2 \delta_{\mu\nu} - \qhat_\mu \qhat_\nu\ ,\nonumber
\\
P^L_{\mu\nu}(q)&=&\qhat_\mu \qhat_\nu\label{eq:proj}
\eea
are the transverse and longitudinal lattice projectors.
In a slight variation on the method of Ref.~\cite{Shintani:2008qe},
we use $\qhat_\mu=(2/a)\sin (q_\mu a/2)$ in \Eq{eq:proj}.
As usual, $q_\mu= (2\pi/L_\mu)n_\mu$ for periodic boundary conditions with period $L_\mu$.

The minimization of $\chi_q^2$
with respect to $\PiLR$ and $\Pi'$ reduces to the simple weighted averages,
\beea
\PiLR(q)  &=& \frac{\sum_{\mu\nu}P^T_{\mu\nu}(q)\Pi^{\textrm{lat}}_{\mu\nu}(q)}{3(\qhat^2)^2},
\\[5pt]
\Pi'(q) &=& \frac{\sum_{\mu\nu}P^L_{\mu\nu}(q)\Pi^{\textrm{lat}}_{\mu\nu}(q)}{(\qhat^2)^2}.
\eea
The success of the decomposition is quantified by calculating
\bee
\Delta(q)= \sum_{\mu\nu} \qhat_\mu \qhat_\nu \left(\frac{1}{\qhat^2} -
\frac{\qhat_\nu}{\sum_\lambda \qhat_\lambda^3}\right)\Pi_{\mu\nu}^{\textrm{lat}}(q).
\label{eq:Delta}
\ee
If $\Pi_{\mu\nu}^{\textrm{lat}}$ is truly a superposition of pure longitudinal and transverse terms [with lattice projectors (\ref{eq:proj})], then $\Delta=0$. We find that the projection works well in our numerical results, observing that $\Delta \ll \PiLR(q)$ up to $q^2 \sim 20$.  For $q^\mu$ oriented along the axes of the lattice, the projection is satisfied exactly, that is, $\Delta = 0$ as expected.

Because of the transverse projector, $\PiLR(q)$ is not defined at $q=0$.
For a finite volume, however, when momenta are discrete, the momentum region around $q=0$ can give a significant contribution to the integral (\ref{eq:CLR1dintegral}).
We handle this region by extrapolation from finite $q$, as will be seen below.

\section{Ensembles and spectra \label{sec:spectra}}

One aim of our study was to compare results for different lattice approximations, and so we created ensembles of configurations for two different values of the NDS coefficient $\gamma$, with values of $(\beta,\kappa)$ chosen to give approximately equal physical scales.
We chose the gauge couplings $(\beta,\gamma)=(7.8,0.125)$ and $(6.0,0.25)$ and a lattice size of $12^3\times 24$ sites.
For each coupling we performed a scan of sea quark $\kappa$ values.
We measured the usual unquenched (i.e., Wilson--clover) spectroscopic quantities---meson masses and decay constants---as well as the Sommer parameter $r_1$ from the static potential.
We then selected one value of $\kappa$ for each gauge coupling to use in the
calculation of $\Pi_{\mu\nu}(q)$.

We list the two ensembles in Table~\ref{tab:ensembles} and the measured observables in Table~\ref{tab:dynspec0}.
As in our earlier work \cite{DeGrand:2015lna}, the quark mass $m_q$ given in Table~\ref{tab:dynspec0} is defined through the axial Ward identity, which relates the divergence of the
axial current $A_\mu^a=\bar \psi \gamma_\mu\gamma_5 (\tau^a/2)\psi$ to the pseudoscalar
density $P^a=\bar \psi \gamma_5 (\tau^a/2)\psi$.
At zero three-momentum we have
\bee
\partial_t \sum_\bx \svev{A_0^a(\bx,t)\co^a} = 2m_q \sum_\bx \svev{ P^a(\bx,t)\co^a}\ ,\quad t>0\ ,
\label{eq:AWI}
\ee
where $\co^a$ is a source at $t=0$, here taken to be a smeared ``Gaussian shell.''
The critical value $\kappa_c$ is determined through the vanishing of the
quark mass $m_q$.
The decay constant  $f_\pi$ has the continuum definition
\bee
 \bra{0} A_\mu^a(x) \ket{\pi^b(p)}
  = ip_\mu f_\pi\, \delta^{ab}\,e^{ipx}.
\label{fpidef}
\ee
In the lattice calculation it is extracted from the ``raw'' value $f_\p^{\text{raw}}$ via \cite{Lepage:1992xa}
\bee
  f_\pi = (1-0.75\kappa/\kappa_c)f_{\pi}^{\text{raw}} \,.
\label{eq:rawfpi}
\ee

The values of $r_1$ are the same in the two ensembles, indicating that the physical lattice size is the same; also the value of $r_1$ shows that the spatial volumes are large enough that confinement physics is undisturbed.
Ensemble~2 is somewhat closer to the massless limit than ensemble~1.
\begin{table}
\begin{ruledtabular}
\begin{tabular}{lcc}
        & Ensemble 1&Ensemble 2\\
\hline
$\beta$ & 6.0      & 7.8      \\
$\gamma$& 1/4      & 1/8      \\
$\kappa$& 0.128    & 0.130    \\
$\kappa_c$&0.1312  &0.1314    \\
Volume  &$12^3\times24$&$12^3\times24$\\
Configurations & 600      & 400      \\
\end{tabular}
\end{ruledtabular}
\caption{Parameters of the two ensembles we analyzed:
gauge coupling $\beta$, NDS coupling $\gamma$, hopping parameter $\kappa$ (and $\kappa_c$ for comparison), lattice volume, and the number of saved configurations.
In both ensembles, configurations were separated by four trajectories of unit length.
In the runs of ensemble 1 the acceptance was 90\%, while in those of ensemble 2 it was 80\%.
We determined $\kappa_c$ by linear extrapolation of $m_q$ (see Table~\ref{tab:dynspec0}) to zero.
\label{tab:ensembles}}
\end{table}
\begin{table}
\begin{ruledtabular}
\begin{tabular}{cll}
        & Ensemble 1&Ensemble 2\\
\hline
$m_q$   & 0.102(1)    & 0.048(1)    \\
$m_{\pi}$& 0.575(2) & 0.392(3) \\
$m_{\rho}$   & 0.750(3) & 0.617(8) \\
$m_{a_1}$   & 1.018(5)  & 0.831(13)\\
$f_{\pi}$& 0.190(1) & 0.140(3) \\
$r_1$   & 3.09(5)  & 3.08(6)
\end{tabular}
\end{ruledtabular}
\caption{Measured observables for the two ensembles described in Table~\ref{tab:ensembles}: Quark mass $m_q$ from the axial Ward identity, meson masses $m_i$, decay constant $f_{\pi}$, and Sommer parameter $r_1$.
The spectra and $f_{\pi}$ are constructed from the same Wilson--clover fermions used in generating the ensembles.
\label{tab:dynspec0}}
\end{table}
The chosen values of $\kappa$ represent the lightest sea quark masses for which the hadronic observables do not show strong effects of finite volume.
Since we take the chiral limit, below, by varying valence masses, we do not venture towards lighter sea quarks.
We omit further discussion of the selection of $\kappa$ values---the analysis is fairly standard and this is not the focus of our work.

We then computed $\Pi_{\mu\nu}(q)$ using valence overlap fermions on the two ensembles, for eight values of the valence quark mass ranging from $m_v=0.01$ to~0.10.  We also supplemented the Wilson--clover observables of Table~\ref{tab:dynspec0} with overlap spectra for these masses.
The spectral calculations were separate from the calculations of $\Pi_{\mu\nu}(q)$ since the inversions used a Coulomb-gauge Gaussian source rather than a point source.
These mixed-action data are collected in Tables~\ref{tab:SU4par6.0v} and~\ref{tab:SU4par7.8v}.

\begin{figure}
\begin{center}
\includegraphics[width=.6\columnwidth,clip]{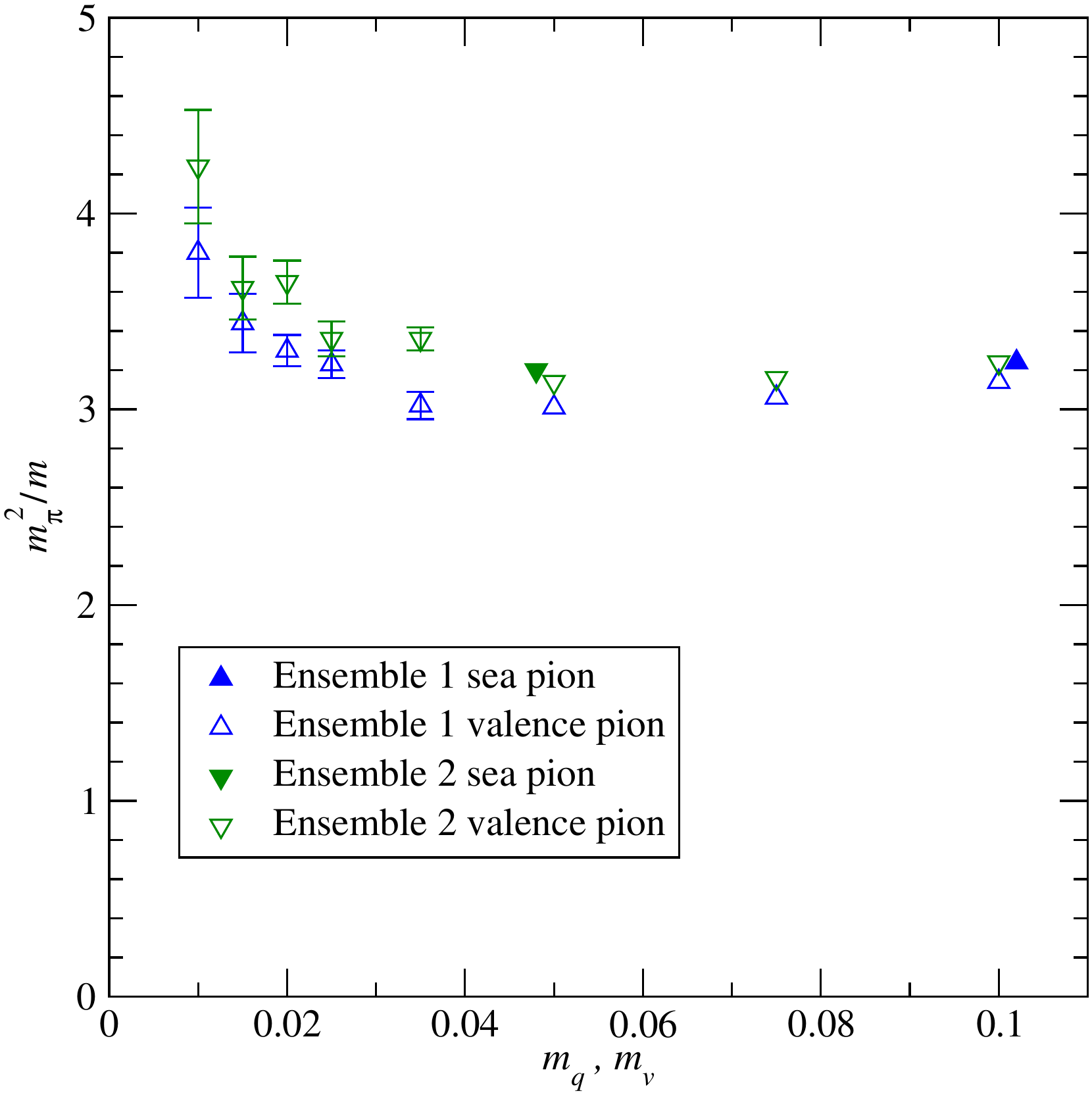}
\end{center}
\caption{
PCAC ratio $m_\pi^2/m$ for valence and sea pions in the two ensembles.
For the valence pions, $m\equiv m_v$, the valence quark mass, while for the sea pions $m\equiv m_q$, the quark mass measured from the axial Ward identity.
Error bars are suppressed if smaller than the plotted symbols.
\label{fig:sqz}}
\end{figure}

We vary the valence quark mass into a regime much lighter than the sea quark, and hence it gives us a handle for checking for chiral logarithms and finite-volume effects.
A plot of $m_\pi^2/m_v$ (see Fig.~\ref{fig:sqz}) shows a common plateau in this ratio for $m_v \agt 0.05$ for the two ensembles, which have a common scale $r_1$.
We note that in each ensemble the sea pion lies comfortably in the plateau furnished by the valence pion.

We can account for the deviations from this plateau using
next-to-leading-order (NLO) mixed-action chiral perturbation theory
($\chi$PT).  The steps needed to obtain the NLO correction for the mass
of the valence pion are similar to those discussed in Appendix \ref{sec:ChiPT}
for the case of the valence pion's decay constant.  A fit to the data in
Fig.~\ref{fig:sqz} with free parameters gives a valid description of the data.

The $p$-regime NLO $\chi$PT formula is comprised of an infinite-volume chiral logarithm
and a finite-volume correction.
Because the parameters of the fit are poorly
determined, we do not know the relative size of these pieces,
nor indeed can we tell whether we are in fact in the $p$-regime.
If we rewrite the $\chi$PT formulas in terms of the valence pion's mass
and decay constant, and then estimate them using the spectroscopic data,
the result suggests that an appreciable part
of the increase in the pion mass seen in Fig.~\ref{fig:sqz} at small $m_v$
originates in finite-volume effects.
Future studies will require larger volumes to accommodate light sea pions as well as to reduce the squeezing of the valence pions.

\begin{table}
\begin{ruledtabular}
\begin{tabular}{ l l l l l }
$m_v$ & $m_{\pi}$ & $m_\rho$ & $m_{a_1}$ & $f_{\pi}$ \\  \hline
0.100 &   0.560(2) &  0.745(4)  &  1.027(7)  &  0.175(1) \\
0.075 &   0.479(2) &  0.690(5)  &  0.975(7)  &  0.162(1) \\
0.050 &   0.388(3) &  0.636(7)  &  0.921(9)  &  0.149(1) \\
0.035 &   0.325(4) &  0.608(10) &  0.886(10) &  0.140(2) \\
0.025 &   0.284(3) &  0.617(11) &  0.879(15) &  0.135(2) \\
0.020 &   0.257(3) &  0.621(9)  &  0.846(13) &  0.128(2) \\
0.015 &   0.227(5) &  0.629(33) & 0.853(18) &  0.127(2) \\
0.010 &   0.195(6) &  0.617(12) &  0.818(16) &  0.120(3) \\
\end{tabular}
\end{ruledtabular}
\caption{
Valence overlap spectra for ensemble~1.
$m_v$ is the valence quark mass.
Note that the order of the lines, here and below, is from heavy to light valence quarks, towards the chiral limit.  We do not show results for $m_q$, extracted from \Eq{eq:AWI}, because it is equal to $m_v$ to the precision shown.}
\label{tab:SU4par6.0v}
\end{table}

\begin{table}
\begin{ruledtabular}
\begin{tabular}{ l l l l l }
$m_v$ & $m_{\pi}$ & $m_\rho$ & $m_{a_1}$ & $f_{\pi}$ \\  \hline
0.100 &   0.569(2) &  0.763(3)  &  0.957(9)  &  0.173(1) \\
0.075 &   0.487(3) &  0.701(5)  &  0.888(0)  &  0.159(1) \\
0.050 &   0.396(4) &  0.651(6)  &  0.817(3)  &  0.142(1) \\
0.035 &   0.343(3) &  0.620(9)  &  0.775(6)  &  0.130(1) \\
0.025 &   0.290(4) &  0.615(12) & 0.790(16) & 0.120(2) \\
0.020 &   0.270(4) &  0.613(11) &  0.734(23) &  0.113(2) \\
0.015 &   0.233(5) &  0.601(18) &  0.764(20) &  0.107(2) \\
0.010 &   0.206(7) &  0.600(17) &  0.678(35) &  0.099(3) \\
\end{tabular}
\end{ruledtabular}
\caption{
Valence overlap spectra for ensemble~2.
\label{tab:SU4par7.8v}}
\end{table}

\clearpage
\section{$\CLR$ via direct integration \label{sec:CLR}}
\subsection{Ultraviolet cutoff\label{sec:UV}}
As a preliminary step, let us determine the expected
dependence of $\CLR$ on an ultraviolet cutoff $M$.
In infinite volume, the value of $\CLR$
depends on the dynamical infrared scale $\Lambda$ of the theory,%
\footnote{
  Here we are concerned with the dependence of $\CLR$
  on the ultraviolet cutoff, for which it suffices to have a rough idea
  of the dynamical infrared scale.  For a detailed discussion,
  see Sec.~\ref{sec:extrap} below.
}
on the fermion mass $m$, and on $M$.
We consider the operator product expansion
for the two-current correlator, which is, schematically,
\bee
  \P_{XX}(q^2;m) \sim 1+ \frac{m^2}{q^2}
     + \frac{g^2\svev{G_{\m\n}G_{\m\n}}+m\svev{\bj\j}}{q^4}
     + \frac{\Lambda^6}{q^6}+ \cdots\ ,
\label{OPEXX}
\ee
where $XX=VV$ or $AA$.  Each term is to be multiplied
by a coefficient function that depends logarithmically on $q^2$.
Note that the first two terms have a perturbative origin.
The $1/q^6$ term comes from several dimension-6 operators.
In the difference $VV-AA$, the identity term drops out, as do all purely gluonic condensates,  and we have
\bee
  \PiLR(q^2;m) \sim \frac{m^2}{q^2} + \frac{m\Lambda^3}{q^4}
     + \frac{\Lambda^6}{q^6} + \cdots\ ,
\label{OPEPiLR}
\ee
where, for each power of $1/q^2$, we show only the leading dependence
on the fermion mass:
in particular, $\svev{\bj\j} \sim \Lambda^3 + \mathcal{O}(m)$.

Introducing the ultraviolet cutoff via
\bee
  \CLR(m;M) = \int_0^{M^2} dq^2 q^2\, \PiLR(q^2;m) \ ,
\label{Ccontcut}
\ee
we have
\bee
  \frac{\partial \CLR}{\partial M^2} = M^2\, \PiLR(M^2;m) \ .
\label{dCdMsq}
\ee
Upon using \Eq{OPEPiLR} and integrating \Eq{dCdMsq}, we find after taking the
chiral limit that
\bee
  \CLR(0;M) = \lim_{m\to0} \CLR(m;M) \sim  \Lambda^4 + \frac{\Lambda^6}{M^2} \ ,
\label{limC}
\ee
where $\Lambda^4$ is the integration constant.
This shows that the dependence of $\CLR=\lim_{M\to\infty}\CLR(0;M)$
on the ultraviolet cutoff vanishes.
This result could have been anticipated by noting that $\CLR$ is
an order parameter for the spontaneous breaking of chiral symmetry.

In our numerical calculation we use a discretized version of \Eq{Ccontcut}.
Since we use chiral valence fermions, the above conclusion applies also on the lattice.
We impose an upper limit $M$ in the summation over momenta, along the lines of \Eq{Ccontcut}, with $M< \pi/a$.
For non-zero (valence) mass, one expects the dependence of $\CLR(m;M)$ on $M$ to follow from Eqs.~(\ref{OPEPiLR}) and~(\ref{dCdMsq}) and hence to contain a quadratically divergent term $\sim m^2M^2$.
In practice, we have seen very little variation of $\CLR$ with  $M$ once $M$ is sufficiently large.
We offer an explanation of this state of affairs in Sec.~\ref{ssec:CLR} below.

\subsection{Summing $\PiLR(q)$ over the lattice momentum\label{sec:summing}}

The calculation of current--current correlators described above gives us $\PiLR(q_\mu)$ at all 4-momenta (except for $q_\mu=0$) for a finite lattice, at a given valence mass $m_v$, where we now consider $\PiLR$ as a function of $q_\mu$, rather than $q^2$.
We write \Eq{eq:CLR1dintegral} in infinite volume as
\begin{equation}
\CLR=16\pi^2\int\frac{d^4q}{(2\pi)^4}\,\PiLR(q_\mu),
\label{integral}
\end{equation}
and integrate directly in four dimensions.
Having in mind a space-time lattice of dimensions $V=L_s^3\times L_t$,
we define the cells in $q$ space such that the side $\mu$ of a cell has length $2\pi/L_\mu$.
We break the integral (\ref{integral}) into a sum of cell integrals,
\begin{equation}
\CLR=\frac{16\pi^2}V \sum_{q_\mu}\PiLR^d(q_\mu).
\label{summation}
\end{equation}
In \Eq{summation}, $\PiLR^d(q_\mu)$ is a discrete quantity, the average of the continuum quantity over the cell in momentum space centered on $q_\mu$:
\begin{equation}
\PiLR^d(q_\mu)=\prod_\nu\left[\frac{L_\nu}{2\pi}\int_{q_\nu-\pi/L_\nu}^{q_\nu+\pi/L_\nu}dq'_\nu\right]
\PiLR(q'_\mu).
\label{average}
\end{equation}
Where $\PiLR$ is smooth, \Eq{average} can be approximated by replacing
$\PiLR^d(q'_\mu)\to\PiLR(q_\mu)$.
That is, we proceed just to sum in \Eq{summation} the values of $\PiLR$ at the centers of the cells, which are the given lattice data.

The exception is the cell centered on $q_\mu=0$, where \Eq{MHA} tells us that $\PiLR$ has a pole.
We approximate the integrand in \Eq{average} by its pole plus a constant pedestal,
\begin{equation}
\PiLR(q_\mu)\simeq p+\frac{f_\pi^2}{q^2},
\label{pole}
\end{equation}
where we take the value of $f_\pi$ from the spectroscopic data, Tables~\ref{tab:SU4par6.0v} and \ref{tab:SU4par7.8v}.
For this cell, then,
\begin{equation}
\PiLR^d(0)\simeq p+\frac{Af_\pi^2}{16\pi^2}\,L_s^2\ ,
\label{av0}
\end{equation}
where
\begin{equation}
A=\int_{-1}^1d^4x\,\frac1{{\bf x}^2+(x_0/b)^2}\ ,
\end{equation}
and we define $b=L_t/L_s$.
The integral for $A$ is readily evaluated for our lattice shape,
\begin{equation}
A(b=2)\circeq22.5095963\cdots\ .
\end{equation}
Inserting into \Eq{av0} and thence into \Eq{summation}, the contribution of the pole to $\CLR$ comes to
\begin{equation}
\frac{L_s^2}V \times A\,f_\pi^2.
\end{equation}
While other cells' contributions to $\CLR$ go as $1/V$, the pole at $q=0$ gives a term $\propto L_s^2/V$, which grows in relative importance as $L_s$ grows.

The contribution of the pedestal to $\CLR$ is $(16\pi^2/V) p$.
The pedestal $p$ can be estimated from the average over some set of cells surrounding the $q=0$ cell.
Collecting $m$ cells at momenta $\lbrace q^a_{\mu}, a=1,\ldots,m\rbrace$, we approximate from \Eq{pole}
\begin{equation}
\sum_a\PiLR^d(q^a)=mp+f_\pi^2\sum_a\frac1{(q^a)^2},
\end{equation}
and thus
\begin{equation}
p=\frac1m\sum_{a=1}^m\left(\PiLR^d(q^a)-f_\pi^2\frac1{(q^a)^2}\right).
\label{neighbors}
\end{equation}
That is, $p$ is the average of the discrete $\PiLR^d$ in the neighboring cells, minus the pole term on those cells.

For this analysis, we made use of six of the eight valence masses, not including the $m_v = 0.015$ and 0.025 data, which were added only later for use with the fitting method discussed in the next section.  The extrapolations to $m_v=0$ are well constrained without the additional valence masses.
Our results are listed in Table~\ref{tab:Cmv} and plotted in Fig.~\ref{fig:plot45}.

The contribution to $\CLR$ of the cell at $q=0$ ranges from about 8\% of the total for the largest $m_v$ to 20\% for the smallest, in both ensembles.
The error bars stand in the same proportion, so the error in \Eq{av0} is not significant in the overall error in $\CLR$.
This of course may change when the lattice dimensions are changed.

In reaching the results shown in Table~\ref{tab:Cmv}, we included in the pedestal (\ref{neighbors}) only the nearest neighbors of $q=0$, namely, the cells displaced by $\pm1$ on the time axis, which are at $q_4=\pm2\pi/24$.
To vary this procedure we included next-nearest neighbors as well.  This means adding the nearest neighbors in the $\pm x$, $\pm y$, and $\pm z$ directions, as well as the next-nearest neighbors in the time direction (all of these have $|q|=2\pi/12$).
Comparing the two pedestal recipes shows a shift of a few percent in $\CLR(m_v)$ at small masses, causing an upward shift of 4\% and 7\% in the chiral extrapolations of Ensemble 1 and Ensemble 2, respectively.
In all cases these are smaller than the statistical error, so we cannot estimate the systematic error due to this part of the algorithm.

\subsection{Chiral extrapolation \label{sec:extrap}}
\begin{figure}
\begin{center}
\includegraphics[width=.6\columnwidth,clip]{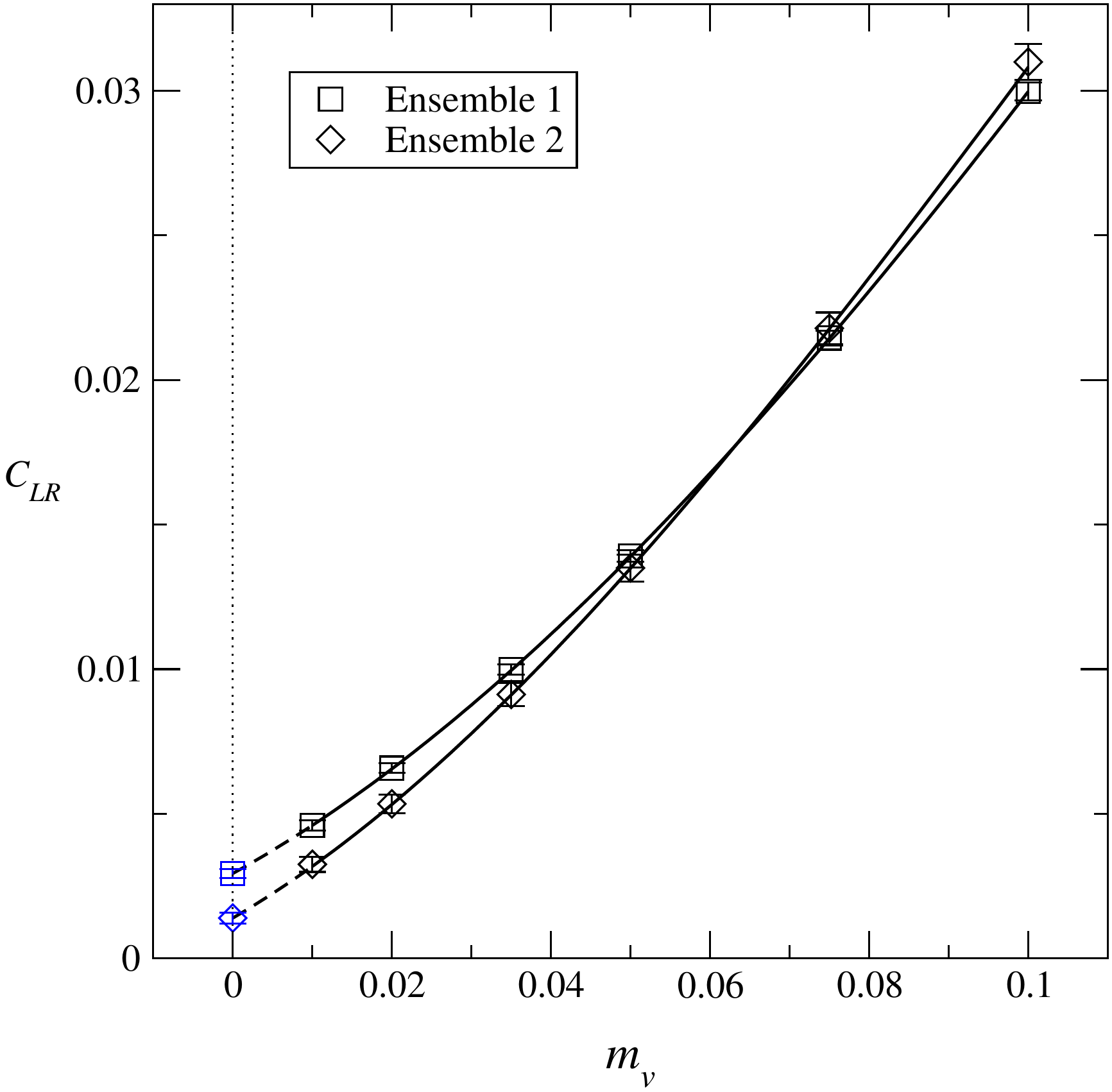}
\end{center}
\caption{
$\CLR(m_v)$ for the two ensembles as listed in Table~\ref{tab:Cmv}.
The extrapolations to $m_v=0$ are cubic fits.
The fit for Ensemble 2 drops the point at $m_v=0.1$.
\label{fig:plot45}}
\end{figure}

$\CLR$ is a quantity that involves an integral over all momenta.
Moreover, in QCD, it is experimentally known that the spectral functions
are dominated by momenta on the order of the $\rho$ and $a_1$ masses.
It is therefore not straightforward to use ChPT
to perform the extrapolation in $m_v$ to $m_v=0$.   At a practical level, while
chiral logarithms may be present (originating from the low-$q^2$ part of the integral),
we find that we are not sensitive to these within our statistics.
Thus, instead, we
adopt a simple power-law fit,
\bee
  \CLR(m_v) = c_0 + c_1 m_v + c_2 m_v^2 + c_3 m_v^3 +  \cdots \,,
\label{fitm}
\ee
giving $c_0$ as the chiral limit.
In practice, we see no advantage to going beyond a cubic.
The result is indicated in Table~\ref{tab:Cmv} and in Fig.~\ref{fig:plot45}.
\begin{table}
\begin{ruledtabular}
\begin{tabular}{ccc}
$m_v$&\multicolumn{2}{c}{$\CLR(m_v)$}\\
\cline{2-3}
& Ensemble 1&Ensemble 2\\
\hline
0.100  & 0.02998(31) & 0.03100(61) \\
0.075  & 0.02145(25) & 0.02179(55) \\
0.050  & 0.01391(20) & 0.01350(47) \\
0.035  & 0.00998(17) & 0.00913(41) \\
0.020  & 0.00658(17) & 0.00534(33) \\
0.010  & 0.00460(17) & 0.00325(26) \\[5pt]
extrapolated $\CLR(0)$  & 0.00293(15) & 0.00151(18) \\
restricted fit       &             & 0.00140(19)
\end{tabular}
\end{ruledtabular}
\caption{$\CLR$ calculated for the two ensembles, at six values of the valence mass
$m_v$, followed by the extrapolation to $m_v=0$ via a cubic [cf.~\Eq{fitm}, and see Fig.~\ref{fig:plot45}]. The fit of Ensemble 1 gives $\chidof=1.0/2$. For Ensemble 2, the cubic fit to all the data points gave $\chidof=7.4/2$.
The last line is the result of a fit that drops the highest-mass point, $m_v=0.100$, giving a satisfactory $\chidof=1.2/1$.
\label{tab:Cmv}}
\end{table}

Strictly speaking, the form of \Eq{fitm} is not adequate for a calculation involving chiral valence propagators evaluated on a Wilson--clover sea.
Zero modes in the valence sector are not suppressed by the Wilson--clover determinant, and hence correlation functions such as $\PiLR$ will contain inverse powers of $m_v$.%
\footnote{This effect was studied in the quenched
approximation in Ref.~\cite{Blum:2000kn}.  For a discussion in the context of mixed action,
see for example Refs.~\cite{Cichy:2010ta,Cichy:2012vg}.}
Because all exact zero modes in any configuration have the same chirality, the structure of $\PiLR$ [\Eq{PiLRdefn}] implies that only one propagator at a time can be saturated by a valence zero mode.
This should lead to a $1/(m_v\sqrt{V})$ divergence at small $m_v$,
where the dependence on the volume $V$ reflects the density
of exact zero modes.
We observe that all our valence chiral extrapolations have good $\chi^2$ without an added
$1/m_v$ term.
We conclude that this effect will show up only at smaller valence masses
than the ones used in this study and/or with improved statistics.

\begin{figure}
\begin{center}
\includegraphics[width=.6\columnwidth,clip]{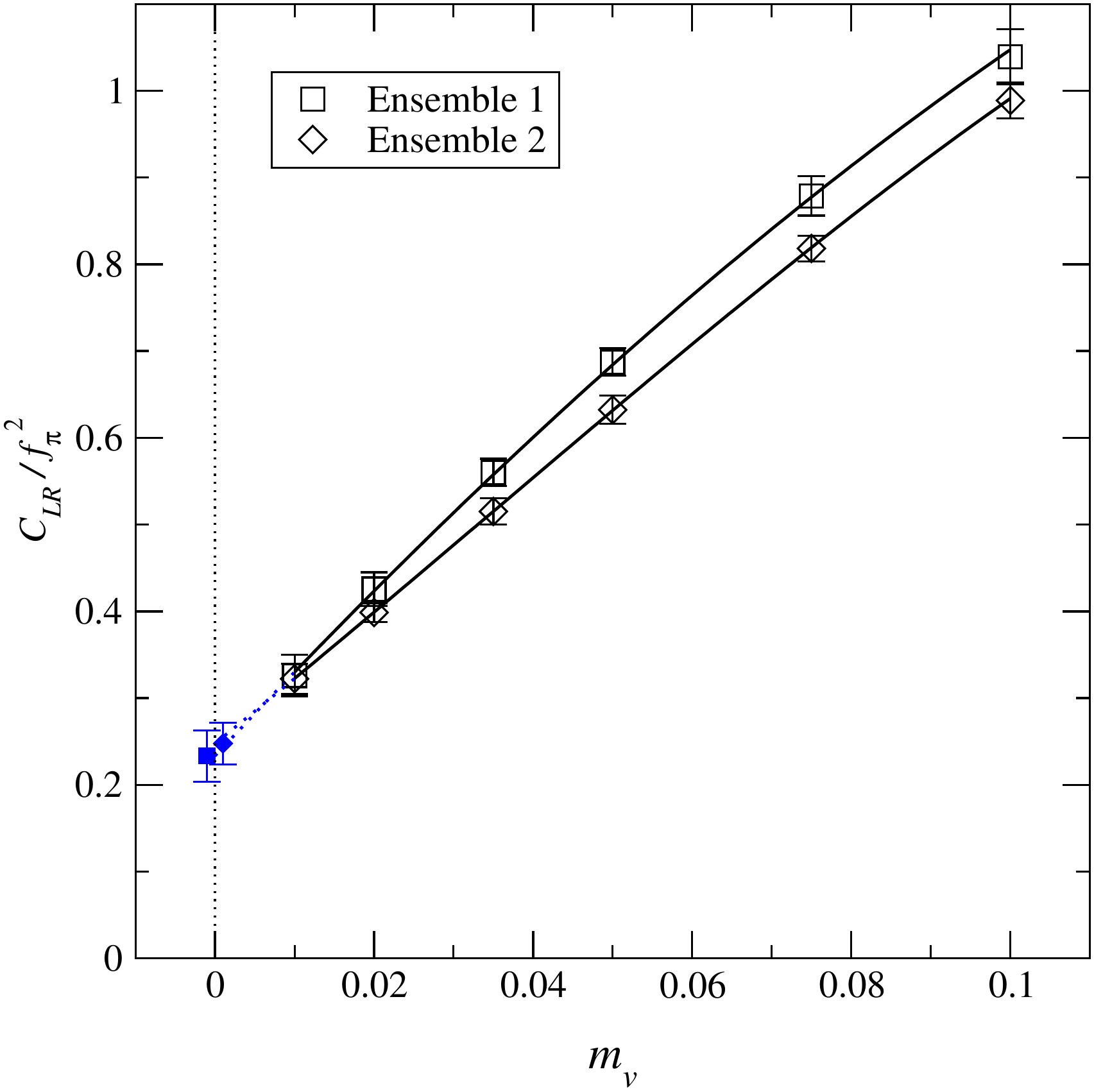}
\end{center}
\caption{
$\CLR(m_v)$ rescaled by $f_\pi^2(m_v)$.
The extrapolations to $m_v=0$ are cubic fits; the extrapolated points are displaced horizontally for clarity.
\label{fig:ratios}}
\end{figure}

The values of $\CLR$ in the chiral limit for the two ensembles differ by a factor of 2, which amounts to more than $6\sigma$, as measured by the statistical errors.
Presumably the origin of this discrepancy is in the difference in lattice actions.
Of all the hadronic observables measurable on the ensembles, the value of $f_\pi$ has entered directly into our determination of $\CLR$, and indeed the MHA, \Eq{CMHA}, gives a direct proportionality between $\CLR$ and $f_\pi^2$.
We see in Tables~\ref{tab:SU4par6.0v} and~\ref{tab:SU4par7.8v} that $f_\pi(m_v)$ of the valence quarks in the two ensembles agrees for large quark masses but diverges for light masses.
We therefore attempt to reconcile the ensembles in their chiral limits by rescaling $\CLR$ by a factor of $f_\pi^2$.
Figure~\ref{fig:ratios}
shows the rescaled quantity $\CLR/f_\pi^2$ for each ensemble.

Equation~(\ref{CMHA}) indicates that the ratio $\CLR/f_\pi^2$ is given by some scale $\Lambda^2$ that stems from the non-Nambu--Goldstone spectrum, that is, from the infrared physics associated with confinement.
As a stand-in for this scale we can take $1/r_1$, since $r_1$ characterizes the confinement distance of the heavy-quark potential.
Referring to Table~\ref{tab:dynspec0} and Fig.~\ref{fig:ratios}, the extrapolations to $m_v=0$ give the dimensionless ratios
\bee
\frac{r_1^2\CLR}{f_\pi^2}=\left\lbrace
\begin{array}{ll}
2.37(24)& \textrm{\qquad Ensemble 1}\\[2pt]
2.21(28)& \textrm{\qquad Ensemble 2}
\end{array}
\right.
\label{eq:CLRr2f2}
\ee
so that perhaps the discrepancy in $f_\pi$ captures the departure of our ensembles from the continuum limit as $m_v\to0$
(we recall that $r_1$ is the same in the two ensembles).
If this is the case, then the path to reconciling the two ensembles might be found in fitting the dependence of the valence $f_\pi$ on the valence mass $m_v$, for both ensembles together, through chiral perturbation theory.
An attempt to do this is detailed in Appendix~\ref{sec:ChiPT}.

Theories with fermions in two-index representations have been studied
in a $1/N_c$ framework \cite{Corrigan:1979xf}, as an alternative to the original
$1/N_c$ expansion that deals with fermions in the fundamental representations.
Either expansion can in principle be applied to QCD, because for $N_c=3$
the fundamental and two-index anti-symmetric representations are the same.
It is therefore interesting to compare the value of the ratio~(\ref{eq:CLRr2f2}) with the QCD value.   Using Eqs.~(\ref{eq:CLR1dintegral}) and (\ref{eq:Das}),
$m_\pi^+-m_\pi^0=4.6$~MeV, and $r_1=0.32$~fm, we have in QCD
\bee
\frac{r_1^2\CLR}{f_\pi^2}=1.9 \qquad\mbox{(QCD)}\ .
\label{eq:QCDvalue}
\ee
The comparison to \Eq{eq:CLRr2f2} may bespeak the validity of large-$N$ arguments.

\section{$\CLR$ via fitting $\PiLR(q^2)$\label{sec:MHA}}

\subsection{Modeling $\PiLR(q^2)$ and the MHA}

An alternative approach to evaluation of the integral (\ref{eq:CLR1dintegral}) is to fit the integrand $q^2 \PiLR(q^2)$ to a smooth function of $q^2$.
If the fit function is extended down to $q^2=0$ then its integral will automatically incorporate the contribution to $\CLR$ from the $q^2\simeq0$ region. We find that a model function inspired by the MHA, \Eq{MHA}, describes our data well when $m_v$ is small enough.  Our fits thus provide a useful cross-check of our direct integration result for $\CLR$, and at the same time testify to the applicability of the MHA.

Equation~(\ref{MHA}) has two poles at $q^2 < 0$ and one at $q^2=0$, so that the function $q^2 \PiLR(q^2)$ in this approximation is finite as $q^2 \rightarrow 0$ and smooth at positive $q^2$.  We can rewrite it as a rational function of $q^2$ with five parameters,
\bee
x\PiLR(x)=a_0\left(\frac{a_1x+1}{a_2x^2+a_3x+1}\right)+b,
\label{eq:fitfn}
\ee
where $x\equiv q^2$.  By comparing \Eq{eq:fitfn} to \Eq{MHA}, we can identify the values of the parameters predicted by the MHA:
\beea
a_0 &=& f_\rho^2 -  f_{a_1}^2, \\
a_1 &=& \frac{f_\rho^2 m_\rho^2 - f_{a_1}^2 m_{a_1}^2}{m_\rho^2 m_{a_1}^2 (f_{\rho}^2 - f_{a_1}^2)}, \nonumber\\
a_2 &=& \frac{1}{m_\rho^2 m_{a_1}^2}, \nonumber\\
a_3 &=& \frac{1}{m_{\rho}^2} + \frac{1}{m_{a_1}^2}, \nonumber\\
b &=& f_{a_1}^2 + f_\pi^2 - f_{\rho}^2.\nonumber
\eea

As a test of the MHA, we use these relations and our fit results to derive the masses and decay constants from $\PiLR(q^2)$. These may then be compared to results for the same quantities from spectroscopy.
We note, however, that there is no a priori reason that the fit parameters of the rational function have to agree with the spectroscopic quantities.
The 3-pole form may be a poor approximation to the multiparticle cuts that characterize the physical current correlation function~\cite{Aubin:2012me}.

The Weinberg sum rules \cite{Weinberg:1967kj} follow in the chiral limit from the requirement that $\PiLR\sim q^{-6}$ as $q^2\to\infty$ [cf.~\Eq{OPEPiLR}].  Combined with the MHA, the sum rules imply the relations
\bee
f_\pi^2=f_\rho^2-f_{a_1}^2
\ee
and
\bee
f_{a_1}^2 m_{a_1}^2=f_\rho^2 m_\rho^2 .
\ee
In terms of our MHA-inspired fit function (\ref{eq:fitfn}), the first condition implies that $a_0 = f_\pi^2$ and $b=0$, while the second condition gives $a_1 = 0$.  At finite mass, we find that our fits to $q^2 \PiLR(q^2)$ require us to keep all five parameters non-zero, but the coefficients $a_1$ and $b$ do appear to be consistent with zero in the chiral limit (see Sec.~\ref{ssec:mha} below).

\subsection{Fitting details}

For each ensemble and at each value of the valence fermion mass $m_v$, we carry out fully correlated, unconstrained least-squares fits to the data for $q^2 \PiLR(q^2)$, making use of the standard \texttt{lsqfit} non-linear fitting package \cite{Lepage:lsqfit,Lepage:2001ym}.  A blocking analysis of our raw data shows evidence for autocorrelations up to a blocking length of 10. In our full analysis we use this blocking length, subsequently treating the blocked data as uncorrelated in Monte Carlo time.

We define $\PiLR(q^2)$ to begin with as an average of $\PiLR(q_\mu)$ over all lattice momenta $q_\mu$ with given length.
We find extremely strong correlations between $\PiLR(q^2)$ values obtained at nearby values of $q^2$.
In order to estimate the data covariance matrix reliably, we reduce the number of $\PiLR$ data points included in the fit.
General arguments show that reliable estimation of an $N \times N$ covariance matrix requires $\mathcal{O}(N^2)$ independent measurements~\cite{Michael:1993yj}.
After blocking to remove autocorrelations, our ensembles consist of 60 (Ensemble 1) and 40 (Ensemble 2) independent configurations, allowing for estimation of covariance matrices of dimension $\sqrt{60}$ and $\sqrt{40}$, respectively.
We therefore limit the number of points in our fits to 6 or~8.%
\footnote{The fit function~(\ref{eq:fitfn}) has 5 parameters, so 6 points still leave us with one degree of freedom.}
Due to the strong correlations in the data, this thinning procedure results in minimal loss of statistical information.

\begin{figure*}
\begin{center}
\includegraphics[width=\textwidth,clip]{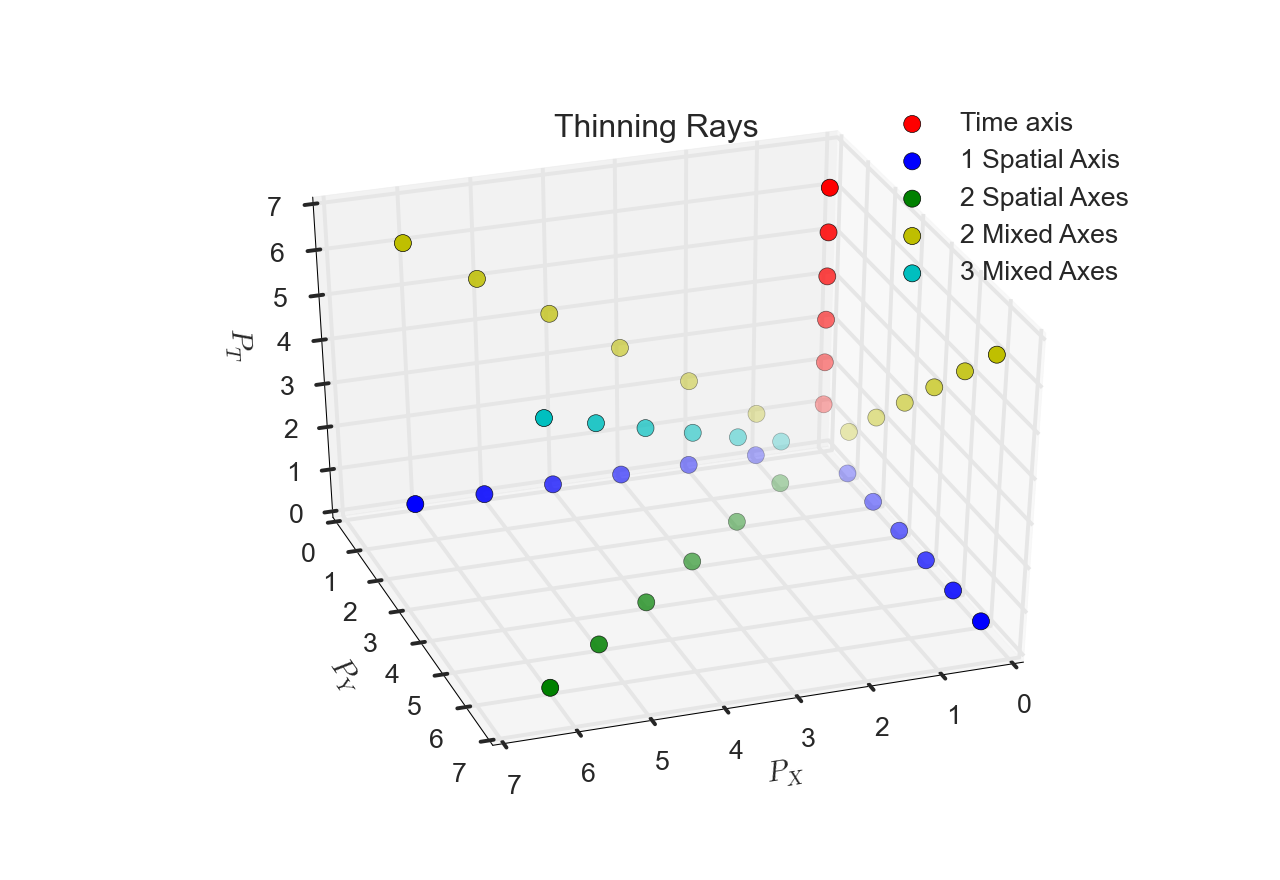}
\end{center}
\caption{A schematic view of the rays defined in Table~\ref{table:thinning-schemes} for a 3-dimensional lattice.
\label{fig:rays}}
\end{figure*}

In order to thin the data, we choose a ray in momentum space (see Table~{\ref{table:thinning-schemes}} and Fig.~\ref{fig:rays}) and use it to select the values of $q^2$.
A ray is defined as an integer multiple $q_\mu=na_\mu$ of a generator vector $a_\mu$, which in turn has one or more nonzero components with the minimal value $2\pi/L_\mu$.
Each momentum vector $q_\mu$ yields a measurement $\PiLR(q^2)$ which is an average of $\PiLR(q^R_\mu)$ over lattice rotations and reflections denoted by $R$; the transformations $\{R\}$ form the subgroup that does {\em not\/} mix time and space axes.
$q_\mu=0$ is always excluded.

Since our lattices have dimension $12^3\times 24$, any ray $q_\mu=na_\mu$ that has a spatial component can reach at most $n=6$ without leaving the first Brillouin zone centered on $q=0$.
Thus we have automatically six values of $\PiLR(q^2)$ to fit once the ray is chosen.
If $a_\mu$ is chosen along the time axis then larger values of $n$ are possible, and here we allow $n$ to reach either 6 or 8 to see whether the results are sensitive to this choice.
Each ray with its maximal $n$ defines a maximum momentum $q_{\rm max}^2$, listed in Table~{\ref{table:thinning-schemes}}.
Varying the ray furnishes an estimate of the systematic error associated with this thinning of the data.

Another source of systematics lies in choosing the definition of the lattice momentum.
In effect, this means choosing between $x=q^2$ and $x=\qhat^2$ (see Sec.~\ref{subsec:qhat}) in defining the fit function~(\ref{eq:fitfn}).
Sec.~\ref{ssec:CLR} below summarizes the variation in our results for $\CLR$ across these choices.

We take the time-axis fit with a $q^2$ cutoff at 6 points as our central fit for each ensemble.  We show the effect of variations on this scheme below.
Even with these thinning schemes, our fits to \Eq{eq:fitfn} fail for $m_v > 0.035$.
We thus report fit results only for $m_v \leq 0.035$, for each ensemble.
We are interested after all in the chiral limit $m_v\to0$.

\begin{table}[h]
\begin{ruledtabular}
\begin{tabular}{l c c}
Ray name & Generator $a_\mu$\footnote{Multiply each component by $(2\pi/L_\mu)$.} & $q_{\rm max}^2\equiv\sum_\mu(6a_\mu)^2$\\
\hline
time axis 	    & $(0,0,0;1) $  & 2.47 \\
1 spatial axis 	& $(1,0,0;0) $	& 9.87 \\
2 spatial axes	& $(1,1,0;0) $	& 19.8 \\
3 spatial axes  & $(1,1,1;0) $	& 29.6\\
2 mixed axes	& $(1,0,0;1) $	& 12.3\\
3 mixed axes	& $(1,1,0;1) $	& 22.2\\
4 mixed axes 	& $(1,1,1;1)$	& 32.1\\
\end{tabular}
\end{ruledtabular}
\caption{Summary of different rays used for thinning in momentum space.
We differentiate between spatial and temporal axes because of the $12^3\times24$ geometry.
See Fig.~\ref{fig:rays}.
}
\label{table:thinning-schemes}
\end{table}

\begin{figure}
\begin{center}
\includegraphics[width=.6\columnwidth,clip]{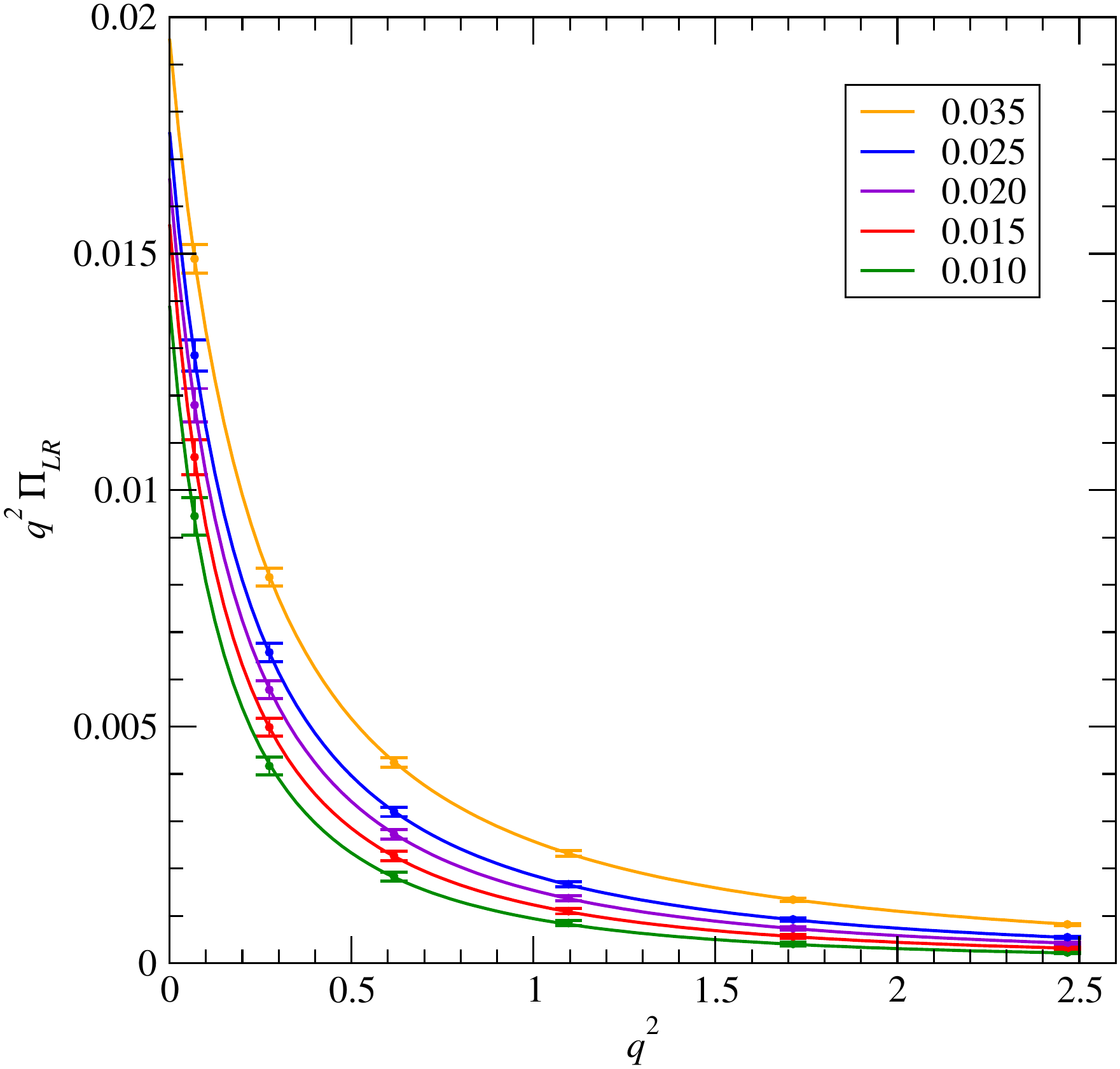}
\end{center}
\caption{Fits to $q^2 \PiLR(q^2)$ with $q^\mu$ oriented in the temporal direction (``time axis'' ray), on Ensemble 1 data at several values of the valence fermion mass $m_v$.  The data and the fit curves are ordered from heaviest to lightest mass, starting at the top of the figure.
\label{fig:mha_fits}}
\end{figure}

\subsection{Testing the validity of the MHA}\label{ssec:mha}

\begin{figure}
\begin{center}
\includegraphics[width=.6\columnwidth,clip]{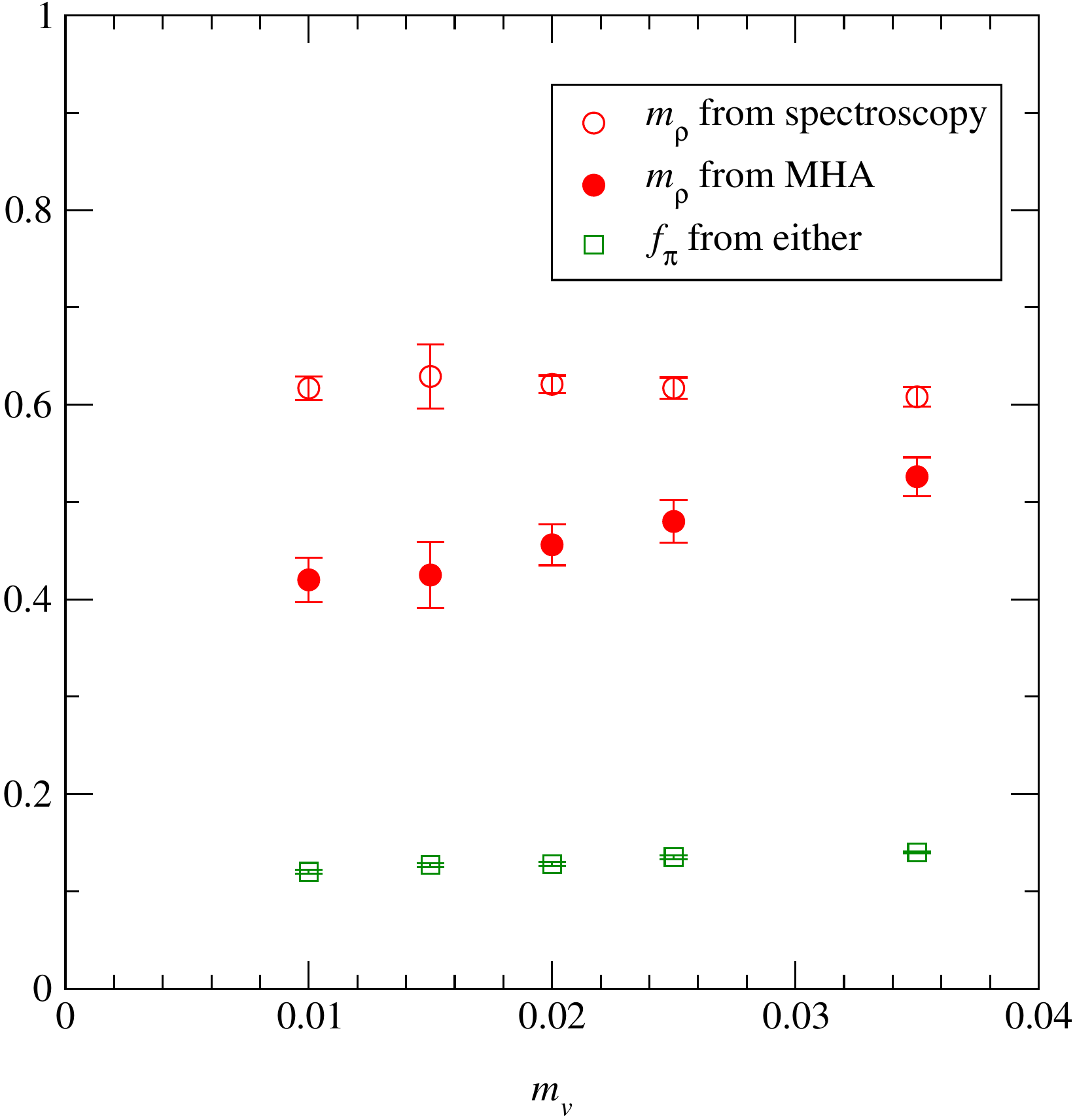}
\end{center}
\caption{
The pion decay constant $f_\pi$ and vector meson mass $m_\rho$ from spectroscopy and from the fits of $q^2\PiLR(q^2)$ to the MHA formula, vs. valence quark mass $m_v$.  The two determinations of $f_\pi$ are indistinguishable.
\label{fig:mha_vs_spec}}
\end{figure}

On the qualitative level, the MHA is a successful formula.
All our fits to \Eq{MHA} [via \Eq{eq:fitfn}] describe the data well up to $m_v=0.035$, with $\chidof \lesssim 1$ in all cases (see Fig.~\ref{fig:mha_fits} for the case of the central fit). On the quantitative level, the MHA is less successful. Table~\ref{table:mha_vs_spec} compares the central-fit results for masses and decay constants to their spectroscopic values. Figure~\ref{fig:mha_vs_spec} presents the same information for
$f_\pi$ and $m_\rho$. The MHA fits are found to reproduce $f_\pi$ beautifully, yielding numbers and uncertainties comparable to direct measurement.%
\footnote{This lends confidence to our use of the spectroscopic value of $f_\pi$ in the direct integration method for estimating $\CLR$.}
The model's other parameters enjoy markedly less success.
In particular, our central-fit results for $m_\rho$ and $f_\rho$ are inconsistent with the  measured spectrum. The axial-vector mass is poorly constrained by the MHA, so that it generally agrees with spectroscopy within large uncertainties.

\begin{figure}
\begin{center}
\includegraphics[width=.6\columnwidth,clip]{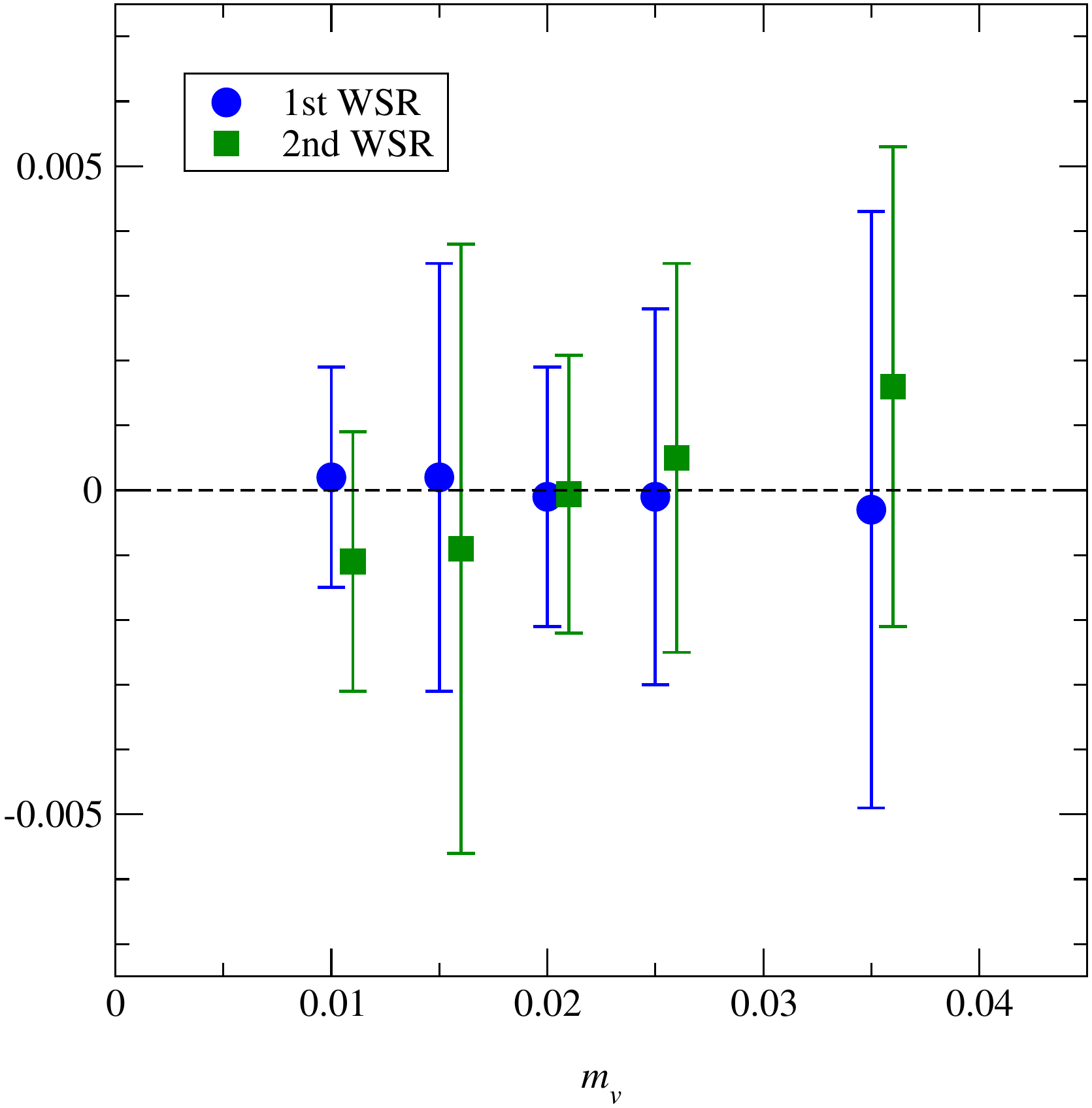}
\end{center}
\caption{
Test of the first and second Weinberg sum rules, based on the quantities WSR1 = $f_\pi^2 -f_\rho^2 +f_{a_1}^2$ (blue circles) and WSR2 = $f_\rho^2 m_\rho^2 - f_{a_1}^2 m_{a_1}^2$ (green squares). Both quantities are expected to vanish in the limit $m_v \rightarrow 0$.
\label{fig:wsr}}
\end{figure}

A complementary test within the MHA is to ask how well the Weinberg sum rules are satisfied by the fit parameters; we expect them to  hold strictly only in the chiral limit.
Figure~\ref{fig:wsr} displays the values of $\text{WSR1} \equiv f_\pi^2 -f_\rho^2 +f_{a_1}^2$ and $\text{WSR2} \equiv f_\rho^2 m_\rho^2 - f_{a_1}^2 m_{a_1}^2$ as functions of valence quark mass; both quantities should be zero in the chiral limit, and this is indeed what we find.
What's more, the central fit gives WSR1 and WSR2 that are consistent with zero (with large error) for all values of $m_v$ up to 0.035.

\begin{table*}
\begin{ruledtabular}
\begin{tabular}{l ll ll ll ll}
\multicolumn{2}{l}{Ensemble 1}& & & & & & \\
$m_v$ & $f_\pi$  & MHA      & $m_\rho$  &  MHA      & $f_\rho$  & MHA       & $m_{a_1}$ & MHA   \\
\hline
0.035 & 0.140(1) & 0.140(2) & 0.608(10) & 0.526(20) & 0.319(13) & 0.160(10) & 0.886(10) & 0.97(14) \\
0.025 & 0.135(2) & 0.133(2) & 0.617(11) & 0.480(22) & 0.360(15) & 0.148(7)  & 0.879(15) & 1.03(20) \\
0.020 & 0.128(2) & 0.129(2) & 0.621(9)  & 0.456(21) & 0.348(3)  & 0.143(5)  & 0.846(13) & 1.09(16) \\
0.015 & 0.127(2) & 0.125(3) & 0.629(33) & 0.425(34) & 0.351(5)  & 0.136(8)  & 0.853(18) & 1.18(46) \\
0.010 & 0.120(2) & 0.118(3) & 0.617(12) & 0.420(23) & 0.347(4)  & 0.129(5)  & 0.818(16) & 1.15(22) \\
\hline \hline
\multicolumn{2}{l}{Ensemble 2}& & & & & & \\
$m_v$ 	& $f_\pi$        	& MHA    		& $m_\rho$       &  MHA  		& $f_\rho$   	& MHA   		& $m_{a_1}$         	& MHA   \\
\hline
0.035 & 0.128(1) & 0.126(2) & 0.620(9)  & 0.560(67) & 0.336(3) & 0.147(24) & 0.775(16) & 0.82(13) \\
0.025 & 0.120(2) & 0.114(2) & 0.615(12) & 0.529(70) & 0.344(4) & 0.130(38) & 0.790(16) & 0.79(15) \\
0.020 & 0.112(2) & 0.106(2) & 0.613(11) & 0.521(59) & 0.340(4) & 0.119(35) & 0.734(23) & 0.71(18) \\
0.015 & 0.107(2) & 0.099(2) & 0.601(18) & 0.477(80) & 0.341(5) & 0.111(33) & 0.764(20) & 0.79(20) \\
0.010 & 0.100(3) & 0.090(3) & 0.600(17) & 0.449(80) & 0.339(4) & 0.098(29) & 0.678(35) & 0.73(33) \\
\end{tabular}
\end{ruledtabular}
\caption{Comparison of MHA best-fit parameters to spectroscopy. The MHA value of $f_\pi$ fits well with spectroscopy, while the $\rho$-meson parameters disagree.  The axial vector mass is generally consistent within large uncertainties.}
\label{table:mha_vs_spec}
\end{table*}

\subsection{The $\CLR$ integral \label{ssec:CLR}}

With the fits to \Eq{eq:fitfn} in hand, we proceed to the one-dimensional integral (\ref{eq:CLR1dintegral}) for $\CLR(m_v)$.
We employ trapezoidal integration directly on the lattice data above $q_{\rm max}^2$, and on the sampled fit function below $q_{\rm max}^2$; carrying out the integral of the fit function analytically yields identical results.
To estimate our statistical errors (for given choice of thinning ray and $q_{\rm max}^2$), we enclose our analysis inside a bootstrap with 500 resamplings.

To determine $\CLR$ in the chiral limit, we must extrapolate from our results at finite fermion mass.
As mentioned above, we were able to carry out the MHA fits only for $m_v\le0.035$.
We again carry out a fully correlated fit, using the bootstrap analysis at each fermion mass $m_v$ with a common set of gauge configuration resamplings to propagate correlations.
A quadratic polynomial in $m_v$ is found to describe the data well; we also carry out a cubic polynomial fit as a consistency check.  As a further test for systematic effects, we repeat the chiral  fits while omitting the lightest mass ($m_v = 0.010$) and omitting the heaviest mass ($m_v = 0.035$).

\begin{figure}
\begin{center}
\includegraphics[width=.6\columnwidth,clip]{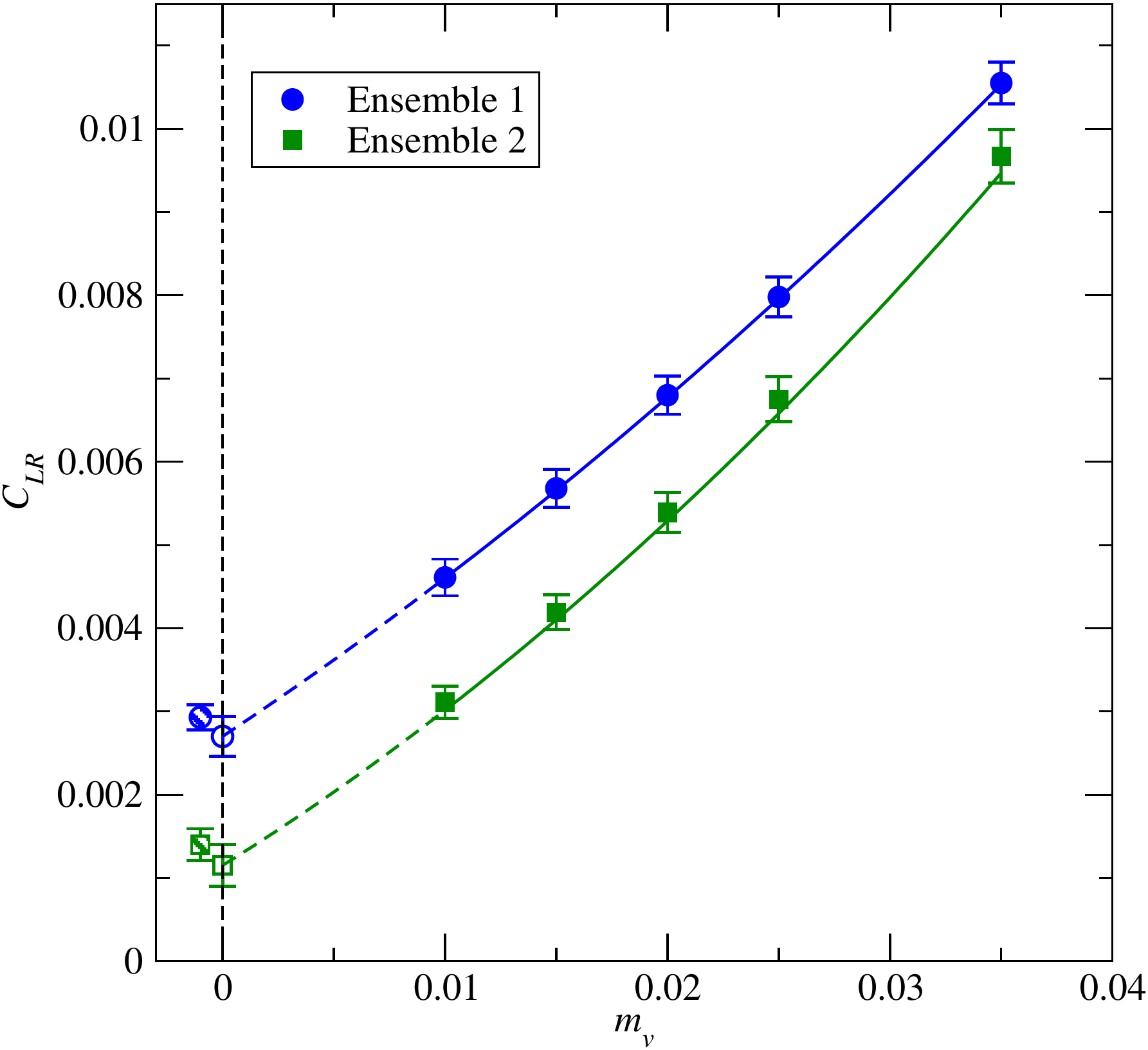}
\end{center}
\caption{
$\CLR(m_v)$ determined with the MHA fits for both ensembles (solid points), and the chiral extrapolations with quadratic polynomials (open points at $m_v=0$).
The extrapolated results show good agreement with those obtained from direct 4d integration (hashed points, displaced horizontally; see Fig.~\ref{fig:plot45}.)
\label{fig:mha_chiral}}
\end{figure}

Figure~\ref{fig:mha_chiral} shows the results of the chiral extrapolations with the quadratic fits.
For comparison, this figure also shows the extrapolated values of $\CLR$ from the direct 4d integration described in Sec.~\ref{sec:CLR}.
The two methods give results that agree within $1\sigma$.

\begin{figure}
\begin{center}
\includegraphics[width=.6\columnwidth,clip]{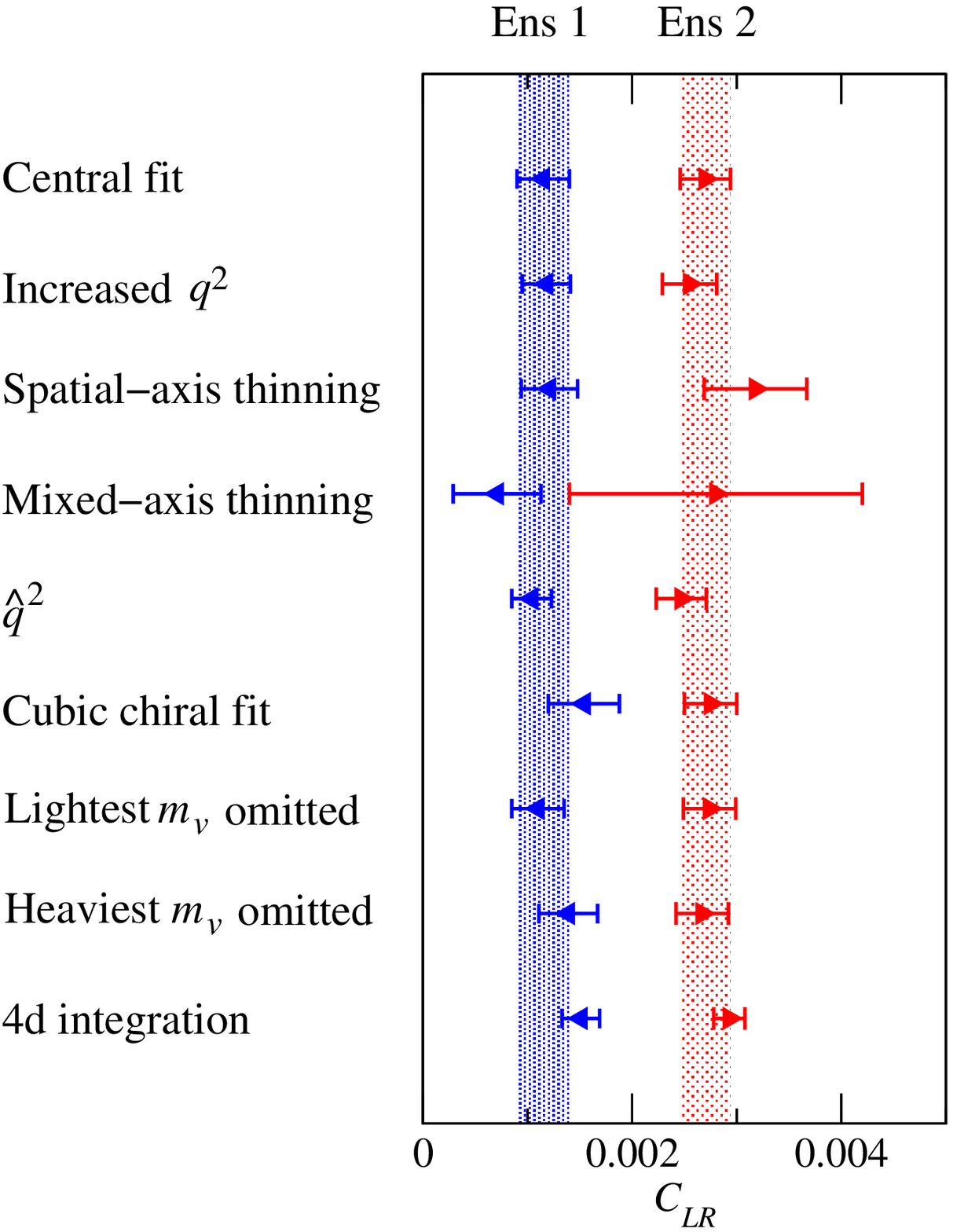}
\end{center}
\caption{
Error stability plot for $\CLR(m_v=0)$ obtained from ensemble 1 (left) and ensemble 2 (right), comparing the central result to other systematic variations as described in the text.  From top to bottom, the variations considered are: increased $q_{\rm max}^2$ to fit to 8 points rather than 6; thinning along a spatial axis; thinning along two mixed axes; using $\hat{q}^2$ instead of $q^2$ to define the lattice momentum; including a cubic term in the chiral extrapolation; and omitting the lightest and heaviest mass points from the chiral extrapolation.  The bottom point is the result of  direct integration from Sec.~\ref{sec:CLR}.
All variations are in agreement with the central fit at roughly $1\sigma$.
\label{fig:mha_stability}}
\end{figure}

Figure~\ref{fig:mha_stability} illustrates the stability of our results for $\CLR$ in the chiral limit across the many systematic variations discussed above. All in all, we considered the following options:
\begin{enumerate}
\item various thinning rays (Table~\ref{table:thinning-schemes}) and a quadratic extrapolation,
\item various thinning rays and a cubic extrapolation,
\item time-axis thinning with a $q^2$ cutoff at 6 or 8 points using a quadratic extrapolation,
\item time-axis thinning with a $q^2$ cutoff at 6 or 8 points using a cubic extrapolation,
\item time-axis thinning with fixed $q^2$ cutoff and a quadratic extrapolation, omitting the heaviest or lightest mass, and
\item all of the above, but using $\hat q^2$ instead of $q^2$ in the fit.
\end{enumerate}
Overall, our results are found to be quite stable, with no significant systematic effects observed. Most importantly, the results present statistically consistent values for $\CLR$ which agree with our original 4d direct integration.

\begin{figure}
\begin{center}
\includegraphics[width=.6\columnwidth,clip]{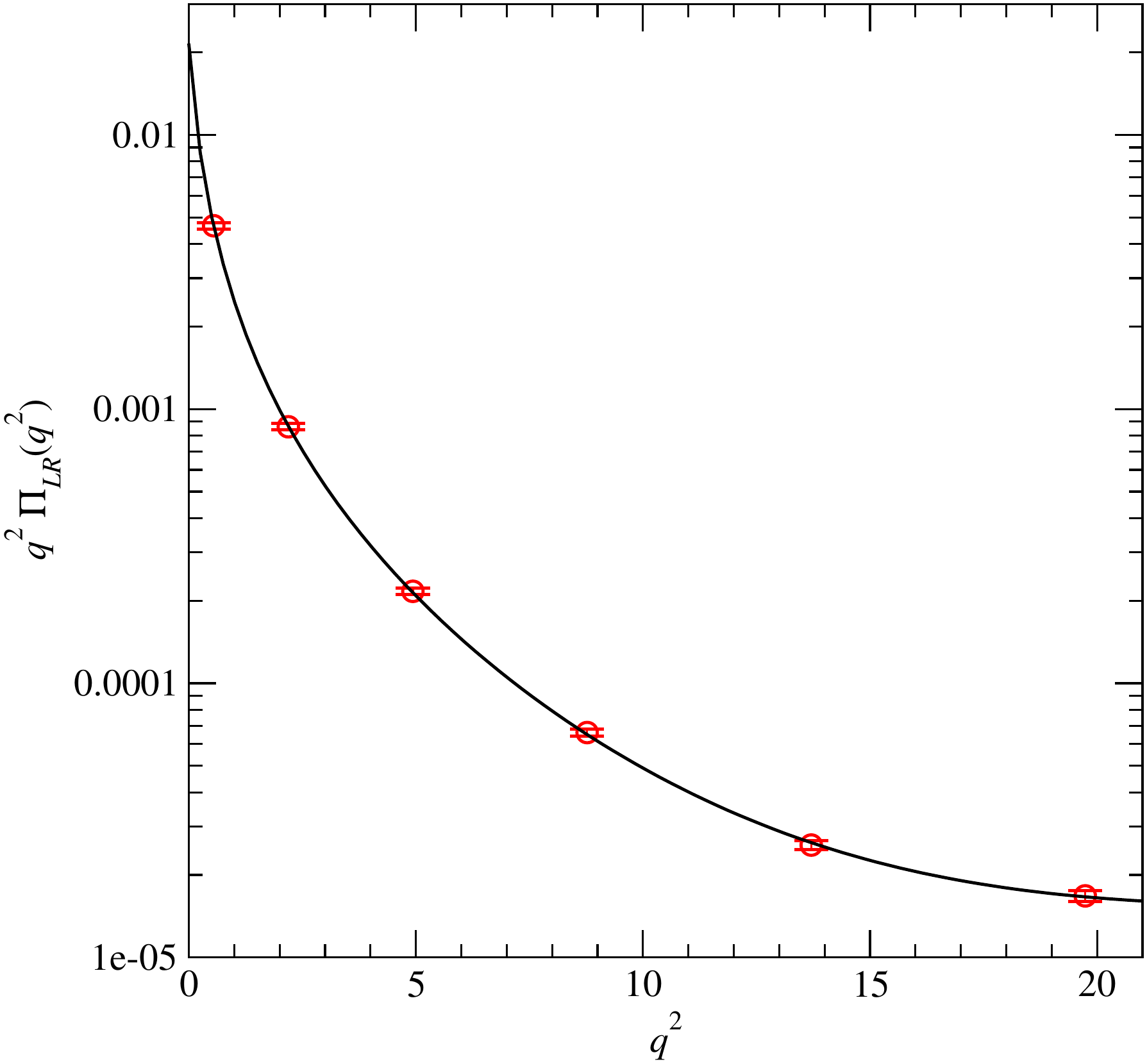}
\end{center}
\caption{
Logarithmic plot of $q^2 \PiLR(q^2)$ with $q^\mu$ oriented along the ``2-spatial-axes'' ray, on Ensemble 1 data with valence fermion mass $m_v=0.035$. \label{fig:spatial2}}
\end{figure}

We believe that the physical reason underlying the stability of our results
across the different analysis methods is the rapid falloff of $\PiLR$
with momentum.  As an illustration,
we show in Fig.~\ref{fig:spatial2} the fit results for $q^2\PiLR(q^2)$
along the 2-spatial-axes ray (see Table~\ref{table:thinning-schemes}), for $m_v=0.035$.
This ray reaches a much larger $q^2$ than the time-axis ray used in Fig.~\ref{fig:mha_fits}.
One can see that the value drops by more than three orders of magnitude
as $q^2$ grows towards $q_{\rm max}^2$.

As noted in Sec.~\ref{sec:UV} [\Eq{OPEPiLR}], the leading term in the OPE of $\PiLR$ is $\sim m_v^2/q^2$, up to logarithms.
The large-$x$ behavior of the MHA fit (5.1) allows us to estimate
this quantity as $b/q^2$.
Defining $c=b/m_v^2$, the fit shown in Fig.~\ref{fig:spatial2} gives $c =0.029(3)$.
Up to a geometrical factor coming from the shape of the
Brillouin zone, $c$ is also the coefficient of the quadratic divergence
$m^2 M^2$ discussed in Sec.~\ref{sec:UV}.  Using the above estimate for $c$
we find that the anticipated variation with the cutoff $M$ is indeed
of roughly the same size as our statistical uncertainties for the entire
range of $m_v$ we have explored.
This explains the insensitivity of $\CLR$ to the ultraviolet cutoff $M$ that we noted in Sec.~\ref{sec:UV}.

Our final results for $\CLR$ from the MHA fitting method are thus
 \bee
\CLR=\left\lbrace
\begin{array}{ll}
0.00270(24)& \textrm{\qquad Ensemble 1}\\[2pt]
0.00115(25)& \textrm{\qquad Ensemble 2}
\end{array}
\right.
\label{eq:CLR_MHA}
\ee
The errors quoted here are the statistical errors on the central fit.
These results are in good agreement with the direct integration results reported in Table~\ref{tab:Cmv}, although not with each other; we refer back to the discussion of Sec.~\ref{sec:extrap} where rescaling by $f_\pi$ and comparison to QCD are considered.


\section{Conclusions \label{sec:conclusions}}

We have carried out a pilot study for the calculation of a radiative contribution to the composite Higgs potential using lattice gauge theory.  We obtained numerical results for an SU(4) gauge theory with two Dirac fermions in the sextet representation.  The electroweak gauge contribution to the potential is determined by the low-energy constant $\CLR$, which can be formulated as an integral over the vacuum polarization function $\PiLR(q^2)$, calculated straightforwardly on the lattice.

We have demonstrated two approaches to the calculation of the integral: a ``direct integration'' technique, which sums over $\PiLR(q^\mu)$ in 4d momentum space, using the value of $f_\pi$ from spectroscopy to account for the pole at $q^2=0$; and a ``rational fit'' technique, which uses a functional form motivated by the minimal hadron approximation  to fit $\PiLR(q^2)$ at low $q^2$.  The $q^2\to0$ behavior of the MHA fit gives an alternative determination of $f_\pi$ which is in good agreement with the spectroscopic value and has similar precision.  The direct integration and the rational fit yield consistent results for $\CLR$.

We have investigated the systematics induced by a number of variations in method, particularly in the context of the MHA fit, including thinning of the data along particular rays in $q^\mu$ space and variations in the chiral extrapolation.  As discussed, these variations show differences of order $1\sigma$ or smaller with the central fit.
Other sources of systematic error, however, have not been studied in this work.  No continuum extrapolation is attempted; neither is the chiral limit of the sea fermions.
Moreover, we are not able to estimate possible finite-volume corrections to our results, even though the squeezing of the valence spectra shows that they must be present.
Within these limitations,
we obtain for the rescaled quantity $r_1^2 \CLR / f_\pi^2 \sim 2.3$, consistent between the two ensembles and with the value of $\CLR$ in QCD inferred from the $m_{\pi^+} - m_{\pi^0}$ mass difference.

The next step in our program is a calculation of $\CLR$ in an SU$(4)$ gauge theory with both fundamental and sextet fermions, for application to the composite Higgs model of Refs.~\cite{Ferretti:2014qta,Ferretti:2016upr}.  Simulation with two Dirac fermions in each representation would be a deviation from the full model, which requires 5 species of Majorana fermions in the sextet representation and 3 Dirac fermions in the fundamental; nonetheless, it should furnish a reasonable approximation while allowing use of the standard hybrid Monte Carlo method.  Calculation of the phase diagram and spectrum of this theory is now underway \cite{toappear}.

\begin{acknowledgments}
This material is based upon work supported by the U.S. Department of
Energy, Office of Science, Office of High Energy Physics, under Award
Number DE-SC0010005 (T.~D. and E.~N.) and Number DE-FG03-92ER40711 (M.~G.).
This work was also supported in part
by the Israel Science Foundation under grant no.~449/13.
Brookhaven National Laboratory is supported by the U.~S.~Department of Energy under contract DE-SC0012704.

Y.~S. thanks San Francisco State University, and
M.~G., Y.~S., and B.~S. thank the University of Colorado, for hospitality.
B.~S. also thanks Prof.~Ting-Wai Chiu for arranging the hospitality of National Taiwan University, where his visit was supported by Taiwan Ministry of Science and Technology under grant no.~MOST 103-2811-M-002-040.
Part of this paper was written while B.~S. and E.~N. were at the Kavli Institute for Theoretical Physics in Santa Barbara, CA; the KITP is supported in part by the National Science Foundation under grant no.~NSF PHY11-25915.

Computations were performed  on clusters of the University of Colorado theory group and of the Tel Aviv University lattice group.
Our computer code is based on version 7 of the publicly available code of the MILC collaboration~\cite{MILC}.
\end{acknowledgments}

\appendix

\section{The NDS term in the action \label{sec:NDS}}
We describe briefly the NDS term \cite{DeGrand:2014rwa} and the reason for its inclusion in the action.

nHYP smearing \cite{Hasenfratz:2001hp} is a three-step process.
In the first step, we average a link $U_{x\rho}$ with two staples in one of the planes containing the link, giving
\bee
  \O = (1-\a_3) U_{x\rho}
  + \frac{\a_3}{2} \left(
    U_{x\x} U_{x+\hat\x,\rho} U^\dagger_{x+\hat\rho,\x}
    + U^\dagger_{x-\hat\x,\x} U_{x-\hat\x,\rho} U_{x-\hat\x+\hat\rho,\x}
  \right) \,.
  \label{NDS:smear}
\ee
We then project $\O$ back to U(4) via
\bee
\bar V_{x,\rho;\x}=\O \bar Q^{-1/2},
\label{NDS:norm}
\ee
where
\bee
\bar Q=\O^\dag\O.
\ee
The following two steps in the smearing average $\bar V$ in the other two planes successively, projecting again each time; we write schematically $U\to \bar V\to \tilde V\to V$, whereupon the fermion action is constructed from the final $V$ links.

Let us focus on the first step.
Since the fermion action contains smeared links $V$, the molecular-dynamics force on the thin link $U_{x\rho}$ is calculated through a chain rule \cite{Hasenfratz:2007rf} that contains derivatives $\partial\bar V/\partial U$.
If one of the plaquettes containing the staples in \Eq{NDS:smear} contains a large field strength (i.e., a dislocation) then $\O$ may be far from unitary, and $\bar Q$ may possess a small eigenvalue.
Then the derivative of \Eq{NDS:norm} can be very large.
Using this force in the time step will introduce a large error in the integration and lead inevitably to rejection of the trajectory.
Since the dislocation is a property of the initial gauge field, it can be very hard for the system to find its way to an acceptable trajectory.

The NDS prescription is designed to keep the system away from regions of phase space where $\bar Q$ has small eigenvalues.
Naturally, it does this for all three steps of the smearing.
The new term in the action takes the simple form,
\bee  S_{\textrm{NDS}} = \frac{1}{2N_c} \sum_x \tr\!\left(
                  \g_1 \sum_{\m}  Q_{x,\m}^{-1}
                + \g_2 \sum_{\m\ne\n} \tilde Q_{x,\m;\n}^{-1}
               + \g_3 \sum_{\rho\ne\x} \bar Q_{x,\rho;\x}^{-1}
      \right) \ ,
\label{NDS:dsa}
\ee
where $\tilde Q$ and $Q$ are the counterparts of $\bar Q$ for the two later smearing steps.
In practice, we take $\gamma_1=\gamma_2=\gamma_3\equiv\gamma$.
This is a pure-gauge term; in weak coupling it shifts the bare coupling according to \Eq{eq:g02}.

\section{The pole at $q^2=0$ \label{sec:pole}}

We rederive here the pion pole contribution to the axial-current
two-point function.
Consider first the massless case.
From the definition of the pion decay constant one might write
\bee
  \svev{A_\m A_\n}(q) = - f_\p^2\, \frac{q_\m q_\n}{q^2} \qquad \text{(incomplete)}.
\label{PiAA}
\ee
Here $A_\m$ is the axial current, $\svev{A_\m A_\n}(q)$ is
the four-dimensional Fourier transform, and we have neglected
the contribution of all other resonances.
It was noted long ago by Brout and Englert \cite{Englert:1964et} that current
conservation requires the presence of a contact term in order that the right-hand side of \Eq{PiAA} be transverse.
With the contact term added, the result is
\bee
  \svev{A_\m A_\n}(q) = f_\p^2 (\delta_{\m\n} - q_\m q_\n/q^2)
  = f_\p^2 P^\perp_{\m\n} \ .
\label{PiAAct}
\ee

Proceeding to the case of a massive pion, we find that the contact term remains unchanged,
while the location of the pole moves:
\beea
  \svev{A_\m A_\n}(q)
  &=& f_\p^2 \left(\delta_{\m\n} - \frac{q_\m q_\n}{q^2+m_\p^2}\right)
\label{PiAAmsv}\\
  &=& f_\p^2 \left(\delta_{\m\n} - \frac{q_\m q_\n}{q^2}\right)
\nonumber\\&&      + f_\p^2 q_\m q_\n
      \left(\frac{1}{q^2} - \frac{1}{q^2+m_\p^2}\right)
\nonumber\\
  &=& P^\perp_{\m\n} f_\p^2 + P^{\|}_{\m\n} \frac{f_\p^2 m_\p^2}{q^2+m_\p^2} \ ,
\nonumber
\eea
where $P^{\|}_{\m\n}=q_\m q_\n/q^2$.
The coefficient
of the transverse projector $P^\perp$ is always $f_\p^2$, irrespective
of $m_\pi$---the dependence on the pion mass is entirely in the longitudinal part.
This result can be traced back to the
derivative coupling of the pion field to the axial current.
It follows that the pion's contribution to $\PiLR(q^2)$ is $f_\p^2/q^2$,
for both massless and massive pions.
The decay constant $f_\p$ depends,
of course, on the quark mass, as does the asymptotic large-$q^2$ behavior.

\section{Mixed action chiral perturbation theory for the valence decay constant \label{sec:ChiPT}}

\def\half{\frac{1}{2}}
\def\irrep{\textit{irrep}}
\def\myskip{\smallskip\noindent}
\def\myskipp{\medskip\noindent}
\def\tA{\tilde{A}}
\def\tB{\tilde{B}}
\def\tC{\tilde{C}}
\def\tD{\tilde{D}}
\def\ha{\hat{a}}
\def\hA{\hat{A}}
\def\hC{\hat{C}}
\def\hD{\hat{D}}
\def\hE{\hat{E}}
\def\hM{\hat{M}}

We saw in Sec.~\ref{sec:extrap} that there is good agreement
between the values of $r_1^2\, \CLR(m_v)/f_\p^2(m_v)$ for the two ensembles
when we extrapolate to the valence chiral limit $m_v\to 0$.
In this appendix we attempt to take this one step further.
We study the measured valence decay constants using mixed-action
chiral perturbation theory ($\chi$PT), to see if we can obtain a prediction
for the continuum limit of the decay constant.
Combined with \Eq{eq:CLRr2f2}, this would allow us to make
a prediction for the continuum limit of $\CLR$ itself.
As it turns out, we cannot obtain a reliable limit for the decay constant from our data. We discuss the possible reasons for this state of affairs.

\subsection{$f_{vv}$ as a function of $m_v$ \label{sec:fitmv}}

Following the conventions of $\chi$PT, in this appendix we refer to the valence
pion mass and decay constant as $M_{vv}(m_v)$ and $f_{vv}(m_v)$, respectively.
Since $r_1$ is practically the same for both ensembles,
instead of working with the dimensionless quantity $r_1 f_{vv}$
we might as well use directly the results reported in lattice units
in Tables~\ref{tab:SU4par6.0v} and~\ref{tab:SU4par7.8v}.

In $N_f=2$ QCD with
Wilson sea quarks and chiral valence quarks, the valence
decay constant $f_{vv}$ is given at the next-to-leading order (NLO) by
\cite{Gasser:1984gg,Bijnens:2009qm,Golterman:2009kw,Rupak:2002sm,Bar:2002nr,Bar:2003mh,Bar:2005tu,Bar:2010qqa}%
\footnote{Most of the $\chi$PT literature uses the convention
$\bra{0} A_\mu^a(x) \ket{\pi^b(p)}   = i\sqrt{2}p_\mu f_\pi\, \delta^{ab}\,e^{ipx}$.
In order to adapt to our convention, \Eq{fpidef},
we make the replacement $f_\pi \to f_\pi/\sqrt{2}$ in the relevant $\chi$PT formulae.
}
\beea
  \frac{f_{vv,NLO}}{f} &=&
  1 - \frac{M_{0,vs}^2}{8\p^2 f^2} \log\left(\frac{M_{0,vs}^2}{\m^2}\right)
\label{fvvnlo}\\&& +\frac{8}{f^2}(L_5 M_{0,vv}^2 + 2L_4 M_{0,ss}^2) + \ha^2\cd\ .
 \nonumber\eea
The leading-order (LO) pion masses are
\beea
  M_{0,vv}^2 &=& 2B m_v \ ,
\label{Mvv}\\
  M_{0,ss}^2 &=& 2(B Z m_s + \ha^2 \cd_{ss}) \ ,
\label{Mss}\\
  M_{0,vs}^2 &=& \half ( M_{0,vv}^2  + M_{0,ss}^2 ) + \hat{a}^2 \cd_{vs} \ .
\label{Mvs}
\eea
Here $f$ and $B$ are the usual parameters of the LO continuum
chiral lagrangian, $f$ being the decay constant in the chiral limit.
$L_4$ and $L_5$ are NLO parameters of continuum $\chi$PT \cite{Gasser:1984gg}.
The parameters $\cd$, $\cd_{ss}$ and $\cd_{vs}$ are linear combinations
of the LO low-energy constants of mixed-action $\chi$PT.
$\ha$ is a rescaled version of the lattice spacing $a$,
roughly, $\ha\sim \Lambda^3 a$, where $\Lambda$ is the confinement scale
\cite{Rupak:2002sm,Bar:2002nr}.
The extra $Z$ factor in \Eq{Mss} arises because our sea-quark's axial current
is not properly normalized, and this renormalization factor has to be
removed from the measured Wilson sea quark's mass.  In other words,
the correct sea-quark mass is $Zm_s$.

In principle
we need to adapt \Eq{fvvnlo} to $N_f=2$ Dirac fermions in a real representation,
for which the coset structure is $\SU(2N_f)/\SO(2N_f)$ rather than the familiar $[\SU(N_f)_L\times \SU(N_f)_R]/\SU(N_f)$ for QCD.
It turns out that the coefficient of the logarithm is unchanged.
Also, while NLO mixed-action results for a real representation are not available,
one can argue that the $O(a^2)$ terms present in
Eqs.~(\ref{fvvnlo}), (\ref{Mss}), and~(\ref{Mvs}) are the most general possible,
and thus, this part of \Eq{fvvnlo} is fine, too.

We will be comparing ensembles with different lattice actions,
as well as different sea masses.  We lump these differences together
by rewriting \Eq{Mvs} as
\bee
  M_{0,vs}^2 = B m_v + A \ ,
\label{MvsA}
\ee
where
\bee
  A = \half M_{0,ss}^2 + \ha^2 \cd_{vs}
  = BZ m_s + \ha^2 (\cd_{ss} + \cd_{vs}) \ .
\label{A}
\ee
Similarly,
\beea
  C &=& 32L_4 (B Z m_s + \ha^2 \cd_{ss})/f^2 + \ha^2 \cd \ ,
\label{C}\\
  D &=& 16L_5 B/f^2 \ .
\label{D}
\eea
Taking the renormalization scale to be $\m=\sqrt{8}\p f$,
Eq.~(\ref{fvvnlo}) takes the form%
\footnote{
  We assume that $L_{4,5}$ have been defined with the same
  renormalization scale.
}
\bee
  f_{vv,NLO} = f \left[
  1 - \left(\frac{A+B m_v}{8\p^2 f^2}\right)
  \log\left(\frac{A+B m_v}{8\p^2 f^2}\right) + C + D m_v \right] \ ,
\label{fvv2}
\ee
In fact, this parametrization suffers from a redundancy.
Eq.~(\ref{fvv2}) is invariant under the reparametrization
\beea
  f &\to& xf \ ,
\label{repar}
\\
  A &\to& xA \ ,
\label{reparra}\nonumber\\
  B &\to& xB \ ,
\label{reparrb}\nonumber\\
  C &\to&  x^{-1}(1-x+C) -(8\p^2 f^2 x)^{-1}A\,\log(x)  \ ,
\label{reparrc}\nonumber\\
  D &\to&  x^{-1}D -(8\p^2 f^2 x)^{-1}B\,\log(x) \ ,
\nonumber\eea
showing that one parameter in \Eq{fvv2} can be eliminated.
Indeed, we can write \Eq{fvv2} in the form
\bee
  f_{vv,NLO} = -(\tA+\tB m_v) \log(\tA+\tB m_v) + \tC
  + \tD m_v \ ,
\label{fvv4}
\ee
by defining
\beea
  \tA &=& A/(8\p^2 f) \ ,
\label{t}\\
  \tB &=& B/(8\p^2 f) \ ,
\nonumber\\
  \tC &=& f(1+C) + \tA\log(f) \ ,
\nonumber\\
  \tD &=& fD + \tB\log(f) \ .
\nonumber
\eea

The fit parameters $\tB$ and $\tD$ depend only
on the low-energy constants of continuum $\chi$PT, so they are universal.
The parameters $\tA$ and $\tC$ depend on the details
of the lattice action, as well as on the sea mass, so they are different
for the two ensembles.  We reserve $\tA$ and $\tC$ for Ensemble 1,
and denote the corresponding parameter of Ensemble 2 by $\tA'$ and $\tC'$.
This makes a total of 6 parameters.

\begin{table*}[t]
\begin{ruledtabular}
\begin{tabular}{ ccccccc }
  $\tB$ & $\tA$ & $\tA'$ & $\tD$ & $\tC$ & $\tC'$ & $\chidof$ \\ \hline
0.358(62) & 0.010(6) & -0.001(3) & -0.324(72) & 0.065(21) & 0.089(15) & 3.1/6
\end{tabular}
\end{ruledtabular}
\caption{Fit parameters for \Eq{fvv4}.
\label{tab:vv}}
\end{table*}

The result of the fit is shown in Table~\ref{tab:vv}.
While the quality of the fit is good, the reparametrization freedom
prevents us from determining $f$.  Several comments are in order.

First, what we are calculating is $f_{vv,NLO}=f_{vv,NLO}(m_s,m_v,\ha)$.
The fit~(\ref{fvv4}) allows us in principle to determine
$f_{vv,NLO}(m_s,0,\ha)$ from the valence chiral extrapolation.
However, this extrapolation is mathematically inconsistent
if $\tA$ or $\tA'$ are negative, because the argument of the logarithm
becomes ill-defined.  We return to this issue below.

Second, additional data at different values of the sea mass $m_s$
and the lattice spacing $\ha$ would in principle allow us
to do also the sea-chiral extrapolation and the continuum
extrapolation, obtaining $f_{vv,NLO}(0,0,0)=f$, the (continuum) decay constant
in the chiral limit.  In other words, the fit does know about $f$
in spite of the reparametrization freedom,
but only through its dependence on all three parameters---$m_s$, $m_v$, and $\ha$.

Third, if we include finite-volume corrections in the $\chi$PT formulae,
the reparametrization freedom disappears.
In principle, this allows us to determine $f$.
In practice, however, the finite-volume NLO correction to $f_{vv}$ is small,
and as a result, $f$ is poorly determined.
This correction is small because the pion that runs in the loop
is a mixed pion [see \Eq{fvvnlo}].  As a result, the finite-volume
correction is not sensitive to the valence pion's mass $M_{vv}$,
but only to the mixed pion's mass $M_{vs}$, which is significantly
larger.

Finally, we note that the reparametrization freedom is quite generic
in Wilson $\chi$PT.  For example, it is also found in the NLO formula
for the dependence of $M^2_{vv,NLO}$ on $m_v$.

\subsection{$f_{vv}$ as a function of $M_{vv}$ \label{sec:fitMvv}}

As an alternative strategy we may fit $f_{vv}$
as a function of the valence pseudoscalar mass.
Thanks to the absence of $B$ at the order we are working,
there is no reparametrization freedom, and it is possible to determine $f$.
The fit form is obtained by replacing $2Bm_v$ by $M^2_{0,vv}$
in \Eq{fvv2}, and then replacing $M^2_{0,vv}$ by the actual measured mass
of the valence pion, $M_{vv}$.  Explicitly,
\bee
  f_{vv,NLO} = f \left[
  1 - (\hA+\hM^2_{vv}) \log(\hA+\hM^2_{vv}) + \hC  + \hD M^2_{vv} \right] \ ,
\label{fMvv}
\ee
where
\beea
  \hM^2_{vv} &=& M^2_{vv}/(16\p^2 f^2) \ ,
\label{fMpar}\\
  \hA &=& A/(8\p^2 f^2) \ ,
\nonumber\\
  \hC &=& C \ ,
\nonumber\\
  \hD &=& D/(2B) \ .
\nonumber
\eea

Since we are now fitting data with errors as a function of data with errors,
the $\c^2$ function takes the form%
\footnote{
  For simplicity we display the $\c^2$ function for uncorrelated data.
  In the actual fit we have taken the cross-correlations into account.
}
\bee
  \c^2 = \sum_{i=1}^n \left[ \frac{(f_i-f(x_i;\a_p))^2}{\delta f_i^2}
  + \frac{(m_i-x_i)^2}{\delta m_i^2} \right] \ .
\label{chisq}
\ee
Here $(f_i,\delta f_i)$ are the measured values and errors of the valence
decay constant,
while $(m_i,\delta m_i)$ are the corresponding data for the valence pion mass
(see Tables~\ref{tab:SU4par6.0v} and~\ref{tab:SU4par7.8v}).
The fit parameters include $\a_p=\{f,\hA,\hA',\hD,\hC,\hC'\}$, as well as
the $x_i$, which are the true values of the valence pion's mass.
Notice that the number of degrees of freedom
in \Eq{chisq} is the same as for the fit~(\ref{fvv2}).

\begin{table*}[t]
\begin{ruledtabular}
\begin{tabular}{ ccccccccc }
data set  & $f$ & $\hA$ & $\hA'$ & $\hD$ & $\hC$ & $\hC'$ &  $\hE$ &
$\chidof$ \\ \hline
1 & 0.07(2) & 0.03(7) & -0.07(3) & 2.2(16) & 0.5(5) & 0.4(3) & -- & 0.16/2  \\
2 & 0.07(2) & 0.03(6) & -0.07(3) & 2.0(12) & 0.4(3) & 0.3(2) & -- & 0.16/2 \\
2 & 0.0346(7) & 0.13(7) & -0.13(5) & 8.3(6) & 1.8(1) & 1.3(1) & 6.5(12) & 0.014/1
\end{tabular}
\end{ruledtabular}
\caption{Fit parameters for \Eq{fMvv}.
Each fit uses 4 values of $m_v$.  For set~1 we used data from Ensembles 1 and 2 for the masses~0.02, 0.035, 0.05 and~0.075;
for set~2, 0.075 was replaced by~0.1.  The first two lines are 6-parameter
fits.  The last line uses set~2 again, but has an added NNLO analytic term
$\hE M_{vv}^4$, for a total of 7 parameters.
\label{tab:fM}}
\end{table*}

Fit results are displayed in Table~\ref{tab:fM}.
In all cases the fit's predictions for the true valence pion's mass, $x_i$,
are consistent with the data, and so we do not display them.
While the fit quality is very good, the errors are rather large.
However, there is a more serious problem.

Mixed-action QCD satisfies the mass inequality \cite{Bar:2010ix}%
\footnote{
  See also Ref.~\cite{Hansen:2011kk}.
}
\bee
  M_{vs} \ge \min(M_{ss},M_{vv}) \ .
\label{ineq}
\ee
Inequality~(\ref{ineq}) must be respected at each order in mixed-action $\chi$PT,
and for any choice of the quark masses
(as long as the masses are not so large that $\chi$PT breaks down).
Constraints on the low-energy constants can be derived
by examining particular limits of the quark masses.
First, we tune the quark masses such that the sea and valence pions
have equal tree-level masses, $M_{0,ss}=M_{0,vv}$.
Using \Eq{Mvs} it follows that we must have
\bee
  \cd_{vs} \ge 0 \ .
\label{Dvs}
\ee
A second inequality follows by considering the double chiral limit,
$(m_s,m_v)\to 0$.  Then $M_{0,vs}\to \ha^2(\cd_{ss} + \cd_{vs})$, and, in order
that $M_{0,vs}^2$ not become negative [in clear violation of \Eq{ineq}],
we must have
\bee
  \cd_{ss} + \cd_{vs} \ge 0 \ .
\label{Dssvs}
\ee
What is important for us is that, using \Eq{A}, the inequality~(\ref{Dvs})
translates into
\bee
  A \ge M_{0,ss}^2/2 \ ,
\label{Amin}
\ee
or, equivalently,
\bee
  \hA \ge M_{0,ss}^2/(16\p^2 f^2) \ .
\label{hatAmin}
\ee
Using the result for $f$ from Table~\ref{tab:fM} (without the $\hat E$ term in the fit),
the ball-park values of this bound are about 0.4 for Ensemble 1 (i.e., $\hat A$),
and 0.2 for Ensemble 2 (i.e., $\hat A'$).  The fit results for $\hA$ and $\hA'$
are in bad disagreement with these bounds.

At a technical level, these difficulties have to do with the large curvature
in the plots of the data for $f_{vv}(M_{vv}^2)$.  Ensemble 2 has the larger curvature,
and, mathematically, the fit makes up for it by demanding a negative $\hA'$.
The bound~(\ref{hatAmin}) is far from satisfied.
Moreover, as the argument of the logarithm is $\hA+\hM^2_{vv}/2$,
the chiral valence extrapolation is mathematically impossible.
In order to avoid this large curvature we would have to work at
smaller (valence and sea) masses, which, in turn, would require larger
lattices to accommodate the pions.


\end{document}